\documentclass[amsmath, pra, twocolumn, showpacs, floatfix]{revtex4-1}
\usepackage{graphicx}
\usepackage{dsfont}

\begin{document}   

\title{Spontaneous emission of a two-level atom with an arbitrarily polarized electric dipole in front of a flat dielectric surface}
 
\author{Fam Le Kien}

\affiliation{Wolfgang Pauli Institute, Oskar Morgensternplatz 1, 1090 Vienna, Austria}

\author{A. Rauschenbeutel} 

\affiliation{Vienna Center for Quantum Science and Technology, Institute of Atomic and Subatomic Physics, Vienna University of Technology, Stadionallee 2, 1020 Vienna, Austria}

\date{\today}

\begin{abstract}
We investigate spontaneous emission of a two-level atom with an arbitrarily polarized electric dipole in front of a flat dielectric surface. 
We treat the general case where the atomic dipole matrix element is a complex vector, that is, the atomic dipole can rotate with time in space.
We calculate the rates of spontaneous emission into evanescent and radiation modes and study the angular densities of the rates in the space of wave vectors for the field modes. We show that, when the ellipticity of the atomic dipole is not zero, the angular density of the spontaneous emission rate of the atom may have different values for modes with opposite in-plane wave vectors. We find that this asymmetry of the angular density of the spontaneous emission rate under central inversion in the space of in-plane wave vectors is a result of spin-orbit coupling of light and occurs when the ellipticity vector of the atomic dipole polarization overlaps with the ellipticity vector of the field mode polarization. Due to the fast decay of the field in the evanescent modes, the difference between the rates of spontaneous emission into evanescent modes with opposite in-plane wave vectors decreases monotonically with increasing distance from the atom to the interface. Due to the oscillatory behavior of the interference between the emitted and reflected fields, the difference between the rates of spontaneous emission into radiation modes with opposite in-plane wave vectors oscillates with increasing distance from the atom to the interface. This difference can be positive or negative, depending on the atom-interface distance, and is zero for the zero distance. 
\end{abstract}

\maketitle

\section{Introduction}
\label{sec:introduction}

The study of individual neutral atoms in the vicinities of material surfaces has a long history 
\cite{Lennard-Jones,Bardeen,Casimir-Polder,Lifshitz,Hoinkes} and has attracted a lot of interest over decades \cite{response,Barton,
Agarwal,Lukosz,conductor,Wylie,Wylie85,dielectric,Courtois,Neugebauer2015,Drexhage,Lima,Oria2006,Fam07,Bennett15,Ducloy14,Jentschura15,Nano-Optics}. The possibility to control and manipulate individual atoms near surfaces can find applications for quantum information \cite{Schlosser,Kuhr,Sackett} and  atom chips \cite{Folman,Eriksson}. Cold atoms can be used as a probe that is very sensitive to surface-induced perturbations \cite{surface probe}. 
Many applications require a deep understanding and an effective control of spontaneous emission of atoms near to material objects. 

It is well known that the spontaneous emission rate of an atom is modified by the presence of an interface \cite{Drexhage,Lukosz,Agarwal,Wylie,Wylie85,conductor,dielectric,Courtois,Fam07,Bennett15,Neugebauer2015,Nano-Optics}. 
Such a modification has been demonstrated experimentally \cite{Drexhage}. A semiclassical approach to the problem of surface-modified radiative properties has been presented \cite{Lukosz}. A quantum-mechanical linear-response formalism has been developed for an atom close to an arbitrary interface \cite{Agarwal,Wylie,Wylie85}.  
An alternative approach based on mode expansion has been used for an atom near a perfect conductor \cite{conductor}. The Green function approach has been applied to a multilayered dielectric \cite{dielectric}. A quantum treatment for the internal dynamics of a multilevel atom near a multilayered dielectric medium has been performed \cite{Courtois}. Spontaneous radiative decay of translational levels of an atom in front of a semi-infinite dielectric has been studied \cite{Fam07}. In the previous treatments \cite{Lukosz,Agarwal,Wylie,Wylie85,conductor,dielectric,Courtois,Fam07,Bennett15,Nano-Optics},
it was assumed that the induced dipole of the atom is linearly polarized, that is, the dipole matrix element vector of the atom is a real vector oriented along a given direction is space. In this condition, the rate of spontaneous emission into evanescent modes is always symmetric with respect to central inversion in the plane of the interface.

In a realistic quantum emitter, the dipole can be elliptically polarized, that is, the dipole matrix element vector can be a complex vector. For example, in an alkali atom, the dipole matrix element vector $\mathbf{d}_{M'M}$ for the transition between the Zeeman levels with the magnetic quantum numbers $M'$ and $M$ is a real vector, aligned along the quantization axis $z$, for the $\pi$ transitions, where $M'=M$, but is a complex vector, lying in the $xy$ plane, for the $\sigma^{\pm}$ transitions, where $M'=M\pm1$. When the dipole matrix element vector is a circularly polarized complex vector, the dipole of the emitter is not aligned along a fixed direction but rotates with time in space. It has recently been shown that spontaneous emission and scattering from an atom with a circular dipole in front of a nanofiber can be asymmetric with respect to the opposite axial propagation directions \cite{Mitsch14b,Petersen14,Scheel15,Sayrin15b,AtomArray,Fam14}. These directional effects are the signatures of spin-orbit coupling of light \cite{Zeldovich,Bliokh review}. They are due to the existence of a nonzero longitudinal component that is in phase quadrature with respect to the radial transverse component of the nanofiber guided field. The possibility of directional emission from an atom into propagating radiation modes of a nanofiber and the possibility of generation of a lateral force on the atom have been pointed out \cite{Scheel15}. 

Spontaneous emission of an atom is similar to the emission of a dipole-like particle.
Spontaneous emission of a two-level atom and radiation of a classical oscillating dipole have identical radiation patterns, identical
rate enhancement factors, and very similar decay rates \cite{Nano-Optics}. A radiating dipole can, in general, oscillate in all three dimensions with relative phases.
Recently, emission of particles with circularly polarized dipoles began to attract much attention \cite{Lee2012,Lin2013,Mueller2013,Zayats2013,Ming2013,Leuchs2014,Banzer15,Zayats,Dogariu}. 
It has been shown that the near-field interference of a circularly polarized dipole coupled to a dielectric or metal leads to unidirectional excitation of guided modes or surface plasmon polariton modes \cite{Lee2012,Lin2013,Mueller2013,Zayats2013,Ming2013,Leuchs2014,Banzer15}. This effect has been experimentally demonstrated by shining circularly polarized light onto a nanoslit \cite{Lee2012,Zayats2013} or closely spaced subwavelength apertures \cite{Lin2013} in a metal film and by exciting a nanoparticle on a dielectric interface with a tightly focused vector light beam \cite{Leuchs2014,Banzer15}. 
The generation of lateral forces by spin-orbit coupling of light scattered off a particle
at an interface between two dielectric media has been demonstrated \cite{Zayats,Dogariu}.
In order to enhance the selective coupling of light to plasmonic and dielectric waveguides on the nanoscale, a variety of complex nanoantenna designs have been proposed and experimentally demonstrated \cite{Curto,Andryieuski,Knoester,King,Fu,Shegai,Vercruysse,Coenen}. Despite recent interest in spin-orbit coupling of light scattered off particles \cite{Mueller2013,Zayats2013,Ming2013,Leuchs2014,Banzer15,Zayats,Dogariu}, a systematic study of the radiation pattern of a circularly polarized dipole in front of an interface is absent. We note that the theory of Ref.~\cite{Nano-Optics} is valid only for  linearly polarized dipoles and must be modified to be used for circularly polarized dipoles \cite{Leuchs2014,Banzer15}. 

Spontaneous emission of a two-level atom and radiation of a classical oscillating dipole are similar but different phenomena.
A two-level atom is a quantum system. The dipole moment of the atom is coupled to the field parametrically.
Meanwhile, the dipole moment of a classical oscillating dipole is coupled directly to the field. 
A quantum atom does not obey the classical equations of motion when the atomic state is far from the ground state. The initial conditions for spontaneous emission are that the field is in the vacuum state and the atom is in the excited state. The spontaneous emission is initiated by the vacuum field fluctuations. 
The expression for the damping rate of a classical oscillating dipole is different from that for the spontaneous emission rate of a two-level atom. In order to get a full understanding of spontaneous emission, the quantum model must be used.

In view of the recent results and insights, it is necessary to develop a systematic theory for spontaneous emission of a two-level atom with an arbitrarily polarized dipole in front of a flat dielectric surface. We construct such a theory in the present paper. We calculate the rates of spontaneous emission into evanescent and radiation modes, and study the angular densities of the rates in the space of wave vectors for the field modes. We focus on the case where the ellipticity of the atomic dipole is not zero, that is, the case where the dipole of the atom rotates with time in space.

The paper is organized as follows. In Sec.~\ref{sec:model} we describe the model and present the expressions for the modes of the field and for 
the Hamiltonian of the atom-field interaction. In Sec.~\ref{sec:rate} we calculate the rates of spontaneous emission into evanescent and radiation modes, and study the angular densities of the rates in the space of wave vectors. In Sec.~\ref{sec:numerical} we present the results of numerical calculations.
Our conclusions are given in Sec.~\ref{sec:summary}.

\section{Model description}
\label{sec:model}

\begin{figure}[tbh]
\begin{center}
\includegraphics{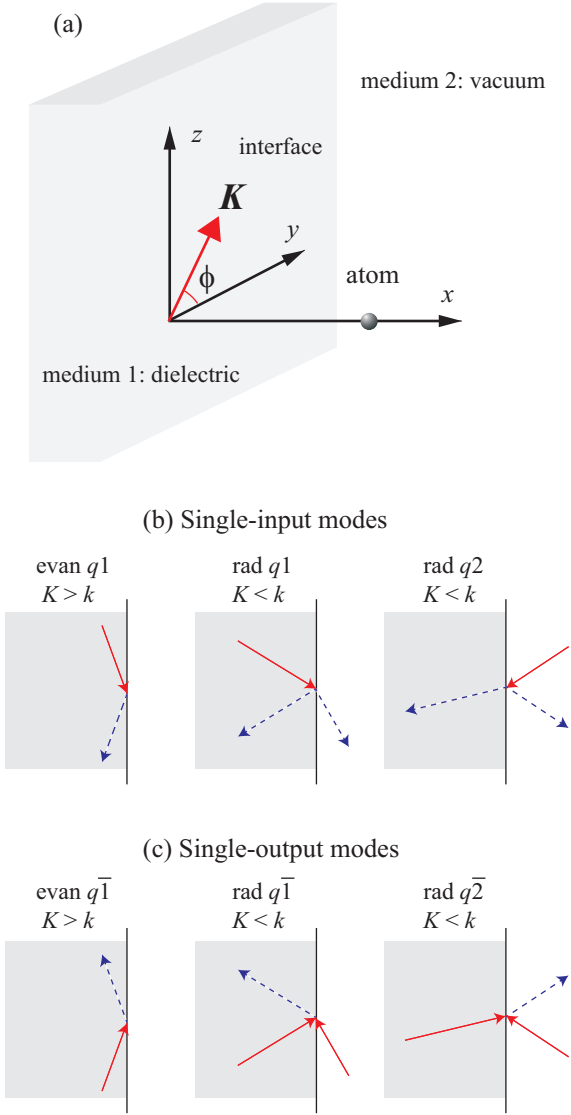}
\end{center}
\caption{(Color online) (a) Atom in front of the flat surface of a semi-infinite dielectric medium.
The half-space $x<0$ is occupied by a dielectric (medium 1).
The half-space $x>0$ is occupied by vacuum (medium 2). The atom lies on the $x$ axis in the half-space $x>0$. 
The axes $y$ and $z$ lie in the interface. The in-plane wave vector $\mathbf{K}$ lies in the interface $yz$ plane.
(b) Representation of single-input modes. (c) Representation of single-output modes.
In (b) and (c), the input and output parts of the modes are shown by the solid red and dashed blue arrows, respectively. 
}
\label{fig1}
\end{figure}

We consider a space with one interface [see Fig.~\ref{fig1}(a)]. We use a Cartesian coordinate system $\{x,y,z\}$. 
The half-space $x<0$ is occupied by a nondispersive nonabsorbing dielectric medium (medium 1).
The half-space $x>0$ is occupied by vacuum (medium 2).
We examine an atom, with an upper energy level $e$ and a lower energy level $g$, located at a fixed point on the $x$ axis in the  half-space $x>0$. The energies of the levels $e$ and $g$ are denoted by $\hbar\omega_e$ and $\hbar\omega_g$, respectively.

We use the formalism of Ref. \cite{Girlanda}
to describe the quantum radiation field in the space with one interface.
We label the modes of the field by the index $\alpha=(\omega\mathbf{K}qj)$, where $\omega$ is the mode frequency, 
$\mathbf{K}=(0,K_y,K_z)$ is the projection of the wave vector onto the dielectric surface $yz$ plane,
$q=s,p$ is the mode polarization index, and $j=1,2$ stands for the medium of the input of the mode. 
For each mode $\alpha=(\omega\mathbf{K}qj)$, the condition $K\leq kn_j$ must be satisfied.
Here, $k=\omega/c$ is the wave number in free space, $n_1>1$ is the refractive index of the dielectric, and $n_2=1$ is the refractive index of the vacuum. We neglect the dependence of the dielectric refractive index $n_1$ on the frequency and the wave number. 

The mode functions are given, for $x<0$, by \cite{Girlanda}
\begin{eqnarray}
\mathbf{U}_{\omega\mathbf{K}s1}(x)&=&\left(e^{i\beta_{1}x}+
e^{-i\beta_{1}x}r_{12}^s\right)\mathbf{s},
\nonumber\\
\mathbf{U}_{\omega\mathbf{K}p1}(x)&=&e^{i\beta_{1}x}\mathbf{p}_{1+}+
e^{-i\beta_{1}x}r_{12}^p\mathbf{p}_{1-},
\nonumber\\
\mathbf{U}_{\omega\mathbf{K}s2}(x)&=&e^{-i\beta_{1}x} t_{21}^s\mathbf{s},
\nonumber\\
\mathbf{U}_{\omega\mathbf{K}p2}(x)&=&e^{-i\beta_{1}x} t_{21}^p\mathbf{p}_{1-},
\label{fl1}
\end{eqnarray} 
and,  for $x>0$, by
\begin{eqnarray}
\mathbf{U}_{\omega\mathbf{K}s1}(x)&=&e^{i\beta_{2}x} t_{12}^s\mathbf{s},
\nonumber\\
\mathbf{U}_{\omega\mathbf{K}p1}(x)&=&e^{i\beta_{2}x} t_{12}^p\mathbf{p}_{2+},
\nonumber\\
\mathbf{U}_{\omega\mathbf{K}s2}(x)&=&\left(e^{-i\beta_{2}x}+
e^{i\beta_{2}x}r_{21}^s\right)\mathbf{s},
\nonumber\\
\mathbf{U}_{\omega\mathbf{K}p2}(x)&=&e^{-i\beta_{2}x}\mathbf{p}_{2-}+
e^{i\beta_{2}x}r_{21}^p\mathbf{p}_{2+}.
\label{fl2}
\end{eqnarray}
In Eqs. (\ref{fl1}) and (\ref{fl2}), 
the quantity $\beta_i=(k^2n_i^2-K^2)^{1/2}$, 
with $\mathrm{Re}\, \beta_i\geq0$ and $\mathrm{Im}\, \beta_i\geq0$,
is the magnitude of the $x$ component of the wave vector in medium $i=1,2$.
The quantities
$r_{ii'}^s=(\beta_{i}-\beta_{i'})/(\beta_{i}+\beta_{i'})$ and
$t_{ii'}^s=2\beta_{i}/(\beta_{i}+\beta_{i'})$
are the reflection and transmission Fresnel coefficients for a TE mode,
while the quantities
$r_{ii'}^p=(\beta_{i}n_{i'}^2-\beta_{i'}n_{i}^2)/(\beta_{i}n_{i'}^2+\beta_{i'}n_{i}^2)$
and $t_{ii'}^p=2n_{i}n_{i'}\beta_{i}/(\beta_{i}n_{i'}^2+\beta_{i'}n_{i}^2)$
are the reflection and transmission Fresnel coefficients for a TM mode.
The vector $\mathbf{s}=[\hat{\mathbf{K}}\times\hat{\mathbf{x}}]$ 
is the polarization vector for the electric field in a TE mode, while the vectors
$\mathbf{p}_{i+}=(K\hat{\mathbf{x}}-\beta_{i}\hat{\mathbf{K}})/kn_{i}$
and $\mathbf{p}_{i-}=(K\hat{\mathbf{x}}+\beta_{i}\hat{\mathbf{K}})/kn_{i}$
are respectively the polarization vectors for the right- and left-moving components of
the electric field in a TM mode in medium $i$. 
Here, the notation $\hat{\mathbf{V}}=\mathbf{V}/V$ stands for the unit vector of an arbitrary vector $\mathbf{V}$, with $V\equiv|\mathbf{V}|=\sqrt{|V_x|^2+|V_y|^2+|V_z|^2}$ being the length of the vector $\mathbf{V}$. 
It is clear from Eqs.~\eqref{fl1} and \eqref{fl2} that each mode
$\alpha=(\omega\mathbf{K}qj)$ has a single input in medium $j$ [see Fig.~\ref{fig1}(b)]. The set of the modes $\alpha$ is a complete and orthogonal basis for the field.

Note that a light beam propagating from the dielectric to the interface may be totally reflected because $n_1>n_2=1$. This phenomenon occurs for the modes $\alpha=(\omega\mathbf{K}qj)$ with $j=1$ and $k<K\leq kn_1$. For such a mode, the magnitude of the $x$ component of the wave vector in medium 2 is $\beta_2=i\sqrt{K^2-k^2}$, an imaginary number. 
This mode does not propagate in the $x$ direction in the vacuum side of the interface but decays exponentially. Such a mode is an evanescent mode. 
We note that, in the case of the $p$ evanescent mode, that is the mode $\alpha=(\omega\mathbf{K}p1)$ with $k<K\leq kn_1$, 
the vector $\mathbf{p}_{2+}$ for the polarization of the field in the half-space $x>0$ is a complex vector. The modes with $0\leq K\leq k$ are called radiation modes.
For convenience, we use the indices $\mu$ and $\nu$ to label the evanescent and radiation modes, respectively, 
that is, we use the notations $\mu=(\omega\mathbf{K}q1)$ with $k<K\leq kn_1$ and $\nu=(\omega\mathbf{K}qj)$ with $0\leq K\leq k$. 

The total quantized electric field is given by \cite{Girlanda}
\begin{equation}
\mathbf{E}(\mathbf{r},t)=i\sum_{\alpha} 
\frac{k}{4\pi}
\sqrt{\frac{\hbar}{\pi\epsilon_0\beta_{j}}}
\;e^{i\mathbf{K}\cdot \mathbf{R}}
\mathbf{U}_{\alpha}(x) a_{\alpha}e^{-i\omega t}
+\mathrm{H.c.},
\label{fl3}
\end{equation}
where $a_{\alpha}$ is the photon annihilation operator for the mode $\alpha$,
$\mathbf{R}=(0,y,z)$ is the projection of the position vector $\mathbf{r}=(x,y,z)$ onto the interface plane, and $\sum_{\alpha}=\sum_{qj}\int_0^{\infty}d\omega
\int_0^{kn_j}K\, dK\int_0^{2\pi}d\phi$
is the generalized summation over the modes. Here, $\phi$ is the azimuthal angle of the
vector $\mathbf{K}$ with respect to the $y$ axis in the $yz$ plane.
The commutation rule for the photon operators is 
$[a_{\alpha},a_{\alpha'}^\dagger]=
\delta(\omega-\omega')\delta(K_y-K'_y)\delta(K_z-K'_z)\delta_{qq'}\delta_{jj'}$.
When dispersion in the region around the frequencies of interest is negligible,
the mode functions $\mathbf{U}_{\alpha}$ satisfy the relation
$\int_{-\infty}^\infty dx\, n^2(x) 
\mathbf{U}_{\omega\mathbf{K}qj}^*(x)\cdot\mathbf{U}_{\omega'\mathbf{K}q'j'}(x)
=2\pi c^2(\beta_{j}/\omega)\delta(\omega-\omega')\delta_{qq'}\delta_{jj'}$.
Here, $n(x)=n_1$ for $x<0$, and $n(x)=n_2$ for $x>0$.
Hence, we can show that the energy of the field is
$\epsilon_0\int d\mathbf{r}\, n^2(x) 
|\mathbf{E}(\mathbf{r})|^2
=\sum_{\alpha}\hbar\omega(
a_{\alpha}^{\dagger}a_{\alpha}+
a_{\alpha}a_{\alpha}^{\dagger})/2$.
Here, $\int d\mathbf{r}=\int_{-\infty}^{\infty} d x\int_{-\infty}^{\infty} d y\int_{-\infty}^{\infty} d z$ is the integral over the whole space.

We now present the Hamiltonian for the atom--field interaction. 
In the dipole and rotating-wave approximations and in the interaction picture, 
the atom--field interaction  Hamiltonian is 
\begin{equation}
H_{\mathrm{int}}=-i\hbar \sum_{\alpha}G_{\alpha}
\sigma^\dagger a_{\alpha}
e^{-i(\omega-\omega_0)t}+\mbox{H.c.},
\label{fl4}
\end{equation}
where $\sigma^\dagger=|e\rangle\langle g|$ describes the atomic transition from the lower level $g$ to the upper level $e$, 
$\omega_0=\omega_e-\omega_g$ is the angular frequency of the transition, and
\begin{equation}
G_{\alpha}=\frac{k}{4\pi\sqrt{\pi\epsilon_0\hbar\beta_{j}}}
e^{i\mathbf{K}\cdot \mathbf{R}}(\mathbf{U}_{\omega\mathbf{K}qj} \cdot\mathbf{d}_{eg})
\label{fl5}
\end{equation}
is the coefficient of coupling between the atom and the mode $\alpha=(\omega\mathbf{K}qj)$. 
In expression (\ref{fl5}),  $\mathbf{d}_{eg}=\langle e|\mathbf{D}|g\rangle$ is the matrix element of the dipole moment operator $\mathbf{D}$ of the atom.
In general, $\mathbf{d}_{eg}$ can be a complex vector.

The time reverse of the mode $\alpha=(\omega\mathbf{K}qj)$ is also a mode of the field.
We introduce the label $\tilde{\alpha}=(\omega,-\mathbf{K},q,\tilde{j})$ for the time reverse of the mode $\alpha=(\omega\mathbf{K}qj)$. The mode function of the mode $\tilde{\alpha}$ is given by $\mathbf{U}_{\tilde{\alpha}}=\mathbf{U}_{\alpha}^*$. 
It is clear that the mode $\tilde{\alpha}$ has a single output coming from the interface into medium $j$ [see Fig.~\ref{fig1}(c)]. 
Like the set of the modes $\alpha$, the set of the modes $\tilde{\alpha}$ is a complete and orthogonal basis for the field.
We can use the basis formed by the modes $\tilde{\alpha}$ instead of the basis formed by the modes $\alpha$.
We note that an evanescent mode $\alpha=(\omega\mathbf{K}qj)$ with $j=1$ and $k<K\leq kn_1$ has a single input and a single output in the dielectric. Thus, we have $(\omega\mathbf{K}qj)=(\omega\mathbf{K}q\tilde{j})$ when $j=1$ and $k<K\leq kn_1$. In other words, there is no difference between single-input evanescent modes and single-output evanescent modes [see the left panels of Figs.~\ref{fig1}(b) and \ref{fig1}(c)].

\section{Spontaneous emission rate}
\label{sec:rate}

We use the mode expansion approach and the Weisskopf--Wigner formalism \cite{Eberly} to derive
the microscopic dynamical equations for spontaneous radiative decay  of 
the atom. We first study the time evolution of an arbitrary atomic operator $\mathcal{O}$. 
The Heisenberg equation for this operator is
\begin{equation}
\dot{\mathcal{O}}=\sum_{\alpha}(G_{\alpha}
[\sigma^\dagger,\mathcal{O}] a_{\alpha}
e^{-i(\omega-\omega_0)t}
+G_{\alpha}^{*}a_{\alpha}^{\dagger}[\mathcal{O},\sigma]
e^{i(\omega-\omega_0)t}).
\label{fl6}
\end{equation}
Meanwhile, the Heisenberg equation for the photon annihilation operator $a_{\alpha}$ is
$\dot{a}_{\alpha}=G_{\alpha}^*\sigma e^{i(\omega-\omega_0)t}$.
Integrating  this equation, we find
\begin{equation}
a_{\alpha}(t)=a_{\alpha}(t_0)+G_{\alpha}^*\int_{t_0}^t dt'\,
\sigma(t')e^{i(\omega-\omega_0)t'}.
\label{fl7}
\end{equation}
Here, $t_0$ is the initial time. For convenience, we take $t_0=0$.

We consider the situation where the field is initially in the vacuum state. 
We assume that the evolution time $t-t_0$ and the characteristic atomic lifetime $\tau$ are 
large as compared to the characteristic optical period $T$. 
Since the continuum of the field modes is broadband, the correlation time of the field bath is short as compared to the atomic lifetime $\tau$. Hence, the Markov approximation $\sigma(t')=\sigma(t)$ can be applied to describe the back
action of the second term in Eq. (\ref{fl7}) on the atom \cite{Eberly}. 
Under the condition $t-t_0\gg T$, 
we calculate the integral with respect to $t'$ in the limit $t-t_0\to\infty$.  
We set aside the imaginary part of the integral, which describes the frequency shift. 
Such a shift is usually small. 
We can effectively account for it  by incorporating it into the atomic frequency and the surface--atom potential. 
With the above approximations, we obtain
\begin{equation}
a_{\alpha}(t)=a_{\alpha}(t_0)+\pi G_{\alpha}^* 
\sigma(t)\delta(\omega-\omega_0).
\label{fl8}
\end{equation}
Inserting Eq. (\ref{fl8}) into Eq. (\ref{fl6}) yields the Heisenberg--Langevin equation
\begin{eqnarray}
\dot{\mathcal{O}}&=&\frac{\gamma}{2}([\sigma^\dagger,\mathcal{O}] \sigma +\sigma^\dagger[\mathcal{O},\sigma])+\xi_{\mathcal{O}}.
\label{fl9}
\end{eqnarray}
Here, 
\begin{equation}
\gamma=2\pi\sum_{\alpha}|G_{\alpha}|^2\delta(\omega-\omega_{0}) 
\label{fl10}
\end{equation}
is the rate of spontaneous emission and $\xi_{\mathcal{O}}$ is the noise operator. 
We emphasize that Eq. (\ref{fl9}) can be applied to any atomic operators.
Due to the presence of the function $\delta(\omega-\omega_{0})$, all the parameters needed for the calculation of the decay rate are to be estimated at the frequency $\omega=\omega_0$.
We will adopt this convention in what follows.

In the half-space $x>0$, where the atom is restricted to, the rate of spontaneous emission $\gamma$ can be decomposed as
\begin{equation}
\gamma=\gamma_{\mathrm{evan}}+\gamma_{\mathrm{rad}}, 
\label{fl11}
\end{equation}
where
\begin{equation}\label{fl12}
\gamma_{\mathrm{evan}}=2\pi\sum_{q=s,p}\int\limits_{k_0}^{k_0n_1}K\, dK\int\limits_0^{2\pi}|G_{\omega_0\mathbf{K}q1}|^2\, d\phi
\end{equation}
is the rate of spontaneous emission into evanescent modes and 
\begin{equation}\label{fl13}
\gamma_{\mathrm{rad}}=2\pi\sum_{q=s,p}\sum_{j=1,2}
\int\limits_{0}^{k_0}K\, dK\int\limits_0^{2\pi}|G_{\omega_0\mathbf{K}qj}|^2\, d\phi
\end{equation}
is the rate of spontaneous emission into radiation modes.

In the particular case where the atom is in free space, that is, where $n_1=n_2=1$, we have $\gamma_{\mathrm{evan}}=0$ and $\gamma=\gamma_{\mathrm{rad}}=\gamma_0$. Here, 
\begin{equation}
\gamma_0=\frac{\omega_0^3d_{eg}^2}{3\pi\epsilon_0\hbar c^3}
\label{fl14}
\end{equation}
is the natural linewidth of the two-level atom \cite{Eberly}.

In the remaining part of this paper, we analyze the consequences of expressions \eqref{fl10}--\eqref{fl13}.
We note that these expressions, apart from a normalization constant equal to $\gamma_0$, can be obtained by using the model 
of an arbitrarily polarized classical oscillating dipole. Consequently, the results of the remaining part of this paper
can be used not only for spontaneous emission of a two-level atom with an arbitrarily polarized dipole but also for the rate enhancement factor and the radiation pattern of an arbitrarily polarized classical oscillating dipole. We emphasize that expression \eqref{fl14} cannot be derived by using the classical formalism. In addition, Eq.~\eqref{fl9} stands for a two-level atom but not for a classical oscillating dipole. This equation describes not only the decay of the atomic level population inversion but also the decay of the atomic coherence.

\subsection{Spontaneous emission into evanescent modes}
\label{subsec:evanescent}

The rate of spontaneous emission from the atom at a position $x>0$ into evanescent modes is
\begin{equation}\label{fl15}
\gamma_{\mathrm{evan}}=\gamma_{\mathrm{evan}}^{s}+\gamma_{\mathrm{evan}}^{p},
\end{equation}
where the notation
\begin{equation}\label{fl16}
\gamma_{\mathrm{evan}}^{q}=2\pi\int\limits_{k_0}^{k_0n_1}K\, dK\int\limits_0^{2\pi}|G_{\omega_0\mathbf{K}q1}|^2\, d\phi
\end{equation}
with $q=s,p$ stands for the rate of spontaneous emission into the $q$-type evanescent modes.

We introduce the notation $\kappa=K/k_0$, where $k_0=\omega_0/c$, for the normalized magnitude of the in-plane component $\mathbf{K}$ of the wave vector. 
In addition, we introduce the notation $\xi=\sqrt{|1-\kappa^2|}$ for the normalized magnitude of the out-of-plane component $\beta_2\hat{\mathbf{x}}$ of the wave vector in the half-space $x>0$. In the case of evanescent modes, we have $\beta_2=ik_0\xi$, $1\le \kappa\le n_1$, and $0\le \xi=\sqrt{\kappa^2-1}\le \sqrt{n_1^2-1}$. In this case, the parameter $\xi$ determines the penetration length $\Lambda=1/k_0\xi$ of the evanescent mode in the half-space $x>0$. We change the integration variable of the first integral in Eq.~\eqref{fl16} from $K$ to $\xi$.
Then, we obtain
\begin{equation}\label{fl17}
\begin{split}
\gamma_{\mathrm{evan}}&=\gamma_0\int\limits_{0}^{\sqrt{n_1^2-1}} \xi d\xi\int\limits_0^{2\pi}F_{\mathrm{evan}}(\xi,\phi)\, d\phi,\\
\gamma_{\mathrm{evan}}^{q}&=\gamma_0\int\limits_{0}^{\sqrt{n_1^2-1}} \xi d\xi\int\limits_0^{2\pi}F_{\mathrm{evan}}^{q}(\xi,\phi)\, d\phi,
\end{split}
\end{equation}
where 
\begin{equation}\label{fl17a}
F_{\mathrm{evan}}=F_{\mathrm{evan}}^{s}+F_{\mathrm{evan}}^{p}, 
\end{equation}
with
\begin{eqnarray}\label{fl18}
F_{\mathrm{evan}}^{s}&=&\frac{3}{4\pi\xi}T_s e^{-2\xi k_0x} 
[|u_y|^2\sin^2\phi+|u_z|^2\cos^2\phi  \nonumber\\
&&\mbox{} -\mathrm{Re\,}(u_y^*u_z)\sin2\phi]         
\end{eqnarray}
and
\begin{eqnarray}\label{fl19}
F_{\mathrm{evan}}^{p}&=&\frac{3}{4\pi\xi}T_p e^{-2\xi k_0x} 
[|u_x|^2(1+\xi^2)+|u_y|^2\xi^2\cos^2\phi \nonumber\\
&&\mbox{} +|u_z|^2\xi^2\sin^2\phi  + \mathrm{Re\,}(u_y^*u_z)\xi^2\sin2\phi \nonumber\\
&&\mbox{} + 2\xi\sqrt{1+\xi^2}\,\mathrm{Im\,}(u_x^*u_y\cos\phi + u_x^*u_z\sin\phi)]. \qquad 
\end{eqnarray}
Here, $u_x$, $u_y$, and $u_z$ are the Cartesian-coordinate components of the unit vector $\mathbf{u}=\mathbf{d}_{eg}/d_{eg}$ for the polarization of the dipole matrix element $\mathbf{d}_{eg}$.
In Eqs.~\eqref{fl18} and \eqref{fl19}, we have introduced the parameters $T_s\equiv (\xi/2\eta)|t_{12}^s|^2$ and $T_p\equiv (\xi/2\eta)|t_{12}^p|^2$, which are proportional to the transmittivity of light coming from medium 1 to medium 2. Here, we have used the notations $t_{12}^s=2\eta/(\eta+i\xi)$, $t_{12}^p=2n_1\eta/(\eta+in_1^2\xi)$, and $\eta\equiv\sqrt{n_1^2-\kappa^2}=\sqrt{n_1^2-1-\xi^2}$.
The explicit expressions for $T_s$ and $T_p$ in terms of $\xi$ are given as
\begin{equation}\label{fl20}
\begin{split}
T_s&=\frac{2\xi\sqrt{n_1^2-1-\xi^2}}{n_1^2-1},\\
T_p&=\frac{2n_{1}^2}{n_1^2-1}\frac{\xi\sqrt{n_1^2-1-\xi^2}}{(n_1^2+1)\xi^2+1}.
\end{split}
\end{equation}

In the half-space $x>0$, the wave vector of an evanescent mode is $(\beta_2,K_y,K_z)$, where $\beta_2=ik_0\xi$.
The parameters $\xi$ and $\kappa=\sqrt{1+\xi^2}$ and the angle $\phi$ characterize the components of the complex wave vector $(\beta_2,K_y,K_z)$ of an evanescent mode in the half-space $x>0$ via the relations $\beta_2/k_0=i\xi$, $K_y/k_0=\kappa_y=\kappa\cos\phi$, and $K_z/k_0=\kappa_z=\kappa\sin\phi$. 

The functions $F_{\mathrm{evan}}^{s}$ and $F_{\mathrm{evan}}^{p}$ are respectively the angular densities of the spontaneous emission rates into the TE evanescent modes $\mu=(\omega_0\mathbf{K}s1)$ and the TM evanescent modes $\mu=(\omega_0\mathbf{K}p1)$, with $k_0< K\leq k_0n_1$, in the wave vector space. The function $F_{\mathrm{evan}}$ is the angular density of the spontaneous emission rate into both $s$ and $p$ types of evanescent modes. In the limit $\kappa\to1$, that is, $K\to k_0$, we have
\begin{eqnarray}\label{fl21}
\lim_{\kappa\to1}F_{\mathrm{evan}}&=&\frac{3}{2\pi}\frac{1}{\sqrt{n_1^2-1}} [n_{1}^2|u_x|^2+|u_y|^2\sin^2\phi\nonumber\\
&&\mbox{}+|u_z|^2\cos^2\phi -\mathrm{Re\,}(u_y^*u_z)\sin2\phi].      
\end{eqnarray}
In the limit $\kappa\to\kappa_{\mathrm{max}}=n_1$, that is, $K\to K_{\mathrm{max}}=k_0n_1$, the rate density $F_{\mathrm{evan}}$ for evanescent modes tends to zero, that is,
we have $\lim_{\kappa\to\kappa_{\mathrm{max}}}F_{\mathrm{evan}}=0$.

In the half-space $x<0$, the wave vector of an evanescent mode is $(\beta_1,K_y,K_z)$, where $\beta_1=k_0\eta$.
Let $\theta$ be the angle between the axis $x$ and the wave vector $(\beta_1,K_y,K_z)$ of the evanescent mode in the dielectric medium.
This angle is determined by the formulas $n_1\sin\theta=\kappa=\sqrt{1+\xi^2}$ and $n_1\cos\theta=-\eta$ for $\theta\in [\pi/2,\pi-\arcsin(1/n_1)]$. We find
$F_{\mathrm{evan}}(\xi,\phi)\xi d\xi d\phi=-P_{\mathrm{evan}}(\theta,\phi)\sin\theta d\theta d\phi$, 
where
\begin{equation}\label{fl21a} 
P_{\mathrm{evan}}=n_1\eta F_{\mathrm{evan}}=-n_1^2\cos\theta F_{\mathrm{evan}}
\end{equation} 
is the angular distribution of spontaneous emission into evanescent modes with respect to the spherical angles $\theta$ and $\phi$. 
The explicit expression for $P_{\mathrm{evan}}$ can be easily obtained by substituting Eq.~\eqref{fl17a} together with Eqs.~\eqref{fl18} and \eqref{fl19} into Eq.~\eqref{fl21a}.
In the particular case where the dipole polarization vector $\mathbf{u}$ is real, this expression reduces to the result
for the far-field limit of the radiation pattern in the forbidden zone of the dielectric \cite{Nano-Optics}.

As already pointed out in the previous section, an evanescent mode $\mu=(\omega_0\mathbf{K}q1)$ with $k_0<K\leq k_0n_1$ has a single input and a single output in the dielectric.
Consequently, there is no difference between single-input evanescent modes and single-output evanescent modes.

The propagation direction of the evanescent mode in the interface plane $yz$ is characterized by the vector $\mathbf{K}=(0,K_y,K_z)$. The transformation $\mathbf{K}\to -\mathbf{K}$ is done by the transformation $\phi\to\phi+\pi$.
We observe that all the terms in expression \eqref{fl18} are associated with the coefficients $\sin^2\phi$, $\cos^2\phi$, and $\sin2\phi$, which do not vary with respect to the transformation $\phi\to\phi+\pi$. Thus, the rate density $F_{\mathrm{evan}}^{s}$ has the same value for the $s$ evanescent modes with the opposite in-plane wave vectors $\mathbf{K}$ and $-\mathbf{K}$. 
Meanwhile, the terms in the last line of expression \eqref{fl19} contain the coefficients $\cos\phi$ and $\sin\phi$,  which change their sign when we replace $\phi$ by $\phi+\pi$. This means that the rate density $F_{\mathrm{evan}}^{p}$ may take different values for the $p$ evanescent modes with  the opposite in-plane wave vectors $\mathbf{K}$ and $-\mathbf{K}$. This asymmetry in spontaneous emission occurs when either $\mathrm{Im\,}(u_x^*u_y)$ or $\mathrm{Im\,}(u_x^*u_z)$ is not zero, that is, when the atomic dipole polarization vector $\mathbf{u}$  is a complex vector and has a nonzero projection onto the axis $x$. The fact that $\mathbf{u}$ is a complex vector means that the direction of the mean dipole $\langle\mathbf{D}(t)\rangle=d_{eg}\langle\mathbf{u}\sigma^\dagger e^{i\omega_0t}+\mathbf{u}^*\sigma e^{-i\omega_0t}\rangle$ of the atom rotates with time in space. The asymmetry of spontaneous emission into evanescent modes with respect to central inversion in the interface plane is a consequence of the interference between the emission from the out-of-plane dipole component $u_x$ and the emission from the in-plane dipole components $u_y$ and $u_z$ where $u_x$ has a phase lag with respect to $u_y$ or $u_z$. When the dipole polarization vector $\mathbf{u}$ is a real vector, the rate density $F_{\mathrm{evan}}$ for evanescent modes is symmetric with respect to central inversion in the interface plane. It is interesting to note that, according to Eq.~\eqref{fl21}, in the limit $\kappa\to1$, 
the rate density $F_{\mathrm{evan}}$ is symmetric with respect to central inversion in the interface plane for an arbitrary dipole polarization vector $\mathbf{u}$.

It is clear from Eqs.~\eqref{fl18} and \eqref{fl19} that the difference $\Delta F_{\mathrm{evan}}\equiv F_{\mathrm{evan}}(\xi,\phi)-F_{\mathrm{evan}}(\xi,\phi+\pi)$ between the rate densities of spontaneous emission into the evanescent modes with the opposite in-plane wave vectors $\mathbf{K}$ and $-\mathbf{K}$ is 
\begin{equation}\label{fl22}
\begin{split}
\Delta F_{\mathrm{evan}}&=\frac{3}{\pi}\sqrt{1+\xi^2} \, T_p e^{-2\xi k_0x}\\
&\quad\times\mathrm{Im\,}(u_x^*u_y\cos\phi + u_x^*u_z\sin\phi).
\end{split}
\end{equation}
We note that the sign (plus or minus) of the rate  density difference $\Delta F_{\mathrm{evan}}$ for evanescent modes depends on 
the dipole polarization vector $\mathbf{u}$ and the azimuthal angle $\phi$ of the in-plane wave vector $\mathbf{K}$ in the $yz$ plane. However, the sign of $\Delta F_{\mathrm{evan}}$ does not depend on the atom-interface distance $x$ and the evanescent-mode penetration parameter $\xi$. When the dipole polarization vector $\mathbf{u}$ is a real vector, the rate density difference for evanescent modes with opposite in-plane wave vectors is $\Delta F_{\mathrm{evan}}=0$.

The asymmetry degree of the angular density $F_{\mathrm{evan}}$ under central inversion in the interface plane is characterized by the factor 
$\zeta_{F_\mathrm{evan}}=\Delta F_{\mathrm{evan}}/F_{\mathrm{evan}}^{\mathrm{sum}}$, where 
$F_{\mathrm{evan}}^{\mathrm{sum}}\equiv F_{\mathrm{evan}}(\xi,\phi)+F_{\mathrm{evan}}(\xi,\phi+\pi)$. It is clear that the asymmetry factor
$\zeta_{F_\mathrm{evan}}$ depends on $\xi$ and $\phi$. However, $\zeta_{F_\mathrm{evan}}$ does not depend on the distance $x$.

We can easily show that
\begin{equation}\label{fl23}
\Delta F_{\mathrm{evan}}=\frac{3}{8\pi\sqrt{n_1^2-1-\xi^2}}[\mathbf{u}^*\times\mathbf{u}]\cdot[\mathbf{U}_{\omega_0\mathbf{K}p1}^{*}\times\mathbf{U}_{\omega_0\mathbf{K}p1}].
\end{equation}
We note that the vector $i[\mathbf{u}^*\times\mathbf{u}]$ is the ellipticity vector of the atomic dipole polarization. Meanwhile,
the vector $-i[\mathbf{U}_{\omega_0\mathbf{K}p1}^{*}\times\mathbf{U}_{\omega_0\mathbf{K}p1}]$ 
is proportional to the ellipticity vector of the local electric polarization of the TM evanescent mode $\mu=(\omega_0\mathbf{K}p1)$ with $K>k_0$ at the position of the atom.
Equation \eqref{fl23} indicates that the difference $\Delta F_{\mathrm{evan}}$ is a result of the overlap between the
ellipticity vector of the atomic dipole polarization and the ellipticity vector of the local electric polarization of the TM evanescent mode $\mu=(\omega_0\mathbf{K}p1)$ with $K>k_0$. The electric part of the other evanescent mode, that is, the TE mode $\mu=(\omega_0\mathbf{K}s1)$ with $K>k_0$, is linearly polarized in the half-space $x>0$. This mode does not contribute to $\Delta F_{\mathrm{evan}}$.

Consider a light field with the electric component $\mathbf{E}=(\boldsymbol{\mathcal{E}}e^{-i\omega t}+\mathrm{c.c.})/2$, where $\boldsymbol{\mathcal{E}}=\mathcal{E}\boldsymbol{\epsilon}$ is the envelope of the positive-frequency component, with $\mathcal{E}$ being the amplitude and $\boldsymbol{\epsilon}$ being the polarization vector. It is known that the local electric spin density $\mathbf{S}$ of the light field is related to the ellipticity vector $-i[\boldsymbol{\epsilon}^{*}\times\boldsymbol{\epsilon}]=\mathrm{Im}\,[\boldsymbol{\epsilon}^{*}\times\boldsymbol{\epsilon}]$ of the local electric polarization via the formula
\begin{equation}\label{fl24}
\mathbf{S}=\frac{\epsilon_0}{4\omega} \mathrm{Im}\big[\boldsymbol{\mathcal{E}}^{*}\times\boldsymbol{\mathcal{E}}\big]=\frac{\epsilon_0}{4\omega} |\mathcal{E}|^2  \mathrm{Im}\big[\boldsymbol{\epsilon}^{*}\times\boldsymbol{\epsilon}\big].
\end{equation}
It follows from Eqs.~\eqref{fl23} and \eqref{fl24} that 
\begin{equation}\label{fl25}
\Delta F_{\mathrm{evan}}=\frac{3\omega_0}{2\pi\epsilon_0\sqrt{n_1^2-1-\xi^2}} i[\mathbf{u}^*\times\mathbf{u}]\cdot\mathbf{S}_{\omega_0\mathbf{K}p1},
\end{equation}
where $\mathbf{S}_{\omega_0\mathbf{K}p1}$ is the local electric spin density of the field in the TM evanescent mode $\mu=(\omega_0\mathbf{K}p1)$ with the in-plane wave number $K>k_0$ 
and the positive-frequency-component envelope $\boldsymbol{\mathcal{E}}_{\omega_0\mathbf{K}p1}=\mathbf{U}_{\omega_0\mathbf{K}p1}$. 
In the half-space $x>0$, where the atom is located, the electric polarization vector of
the TM evanescent mode $\mu=(\omega\mathbf{K}p1)$ with $K>k$ is
\begin{equation}\label{fl26a} 
\boldsymbol{\epsilon}_{\omega\mathbf{K}p1}=\frac{\kappa\hat{\mathbf{x}}-i\xi\hat{\mathbf{K}}}{\sqrt{\kappa^2+\xi^2}}. 
\end{equation}
The ellipticity vector of the electric polarization of the field is found to be
\begin{equation}\label{fl26}
\mathrm{Im}\big[\boldsymbol{\epsilon}_{\omega\mathbf{K}p1}^{*}\times\boldsymbol{\epsilon}_{\omega\mathbf{K}p1}\big]
=2\frac{\xi\sqrt{1+\xi^2}}{1+2\xi^2}[\hat{\mathbf{K}}\times \hat{\mathbf{x}}],
\end{equation}
which leads to the local electric spin density
\begin{equation}\label{fl27}
\mathbf{S}_{\omega\mathbf{K}p1}=\frac{\epsilon_0}{\omega}\frac{2n_{1}^2}{n_1^2-1}\frac{n_1^2-1-\xi^2}{(n_1^2+1)\xi^2+1}\xi\sqrt{1+\xi^2} \, e^{-2\xi kx} [\hat{\mathbf{K}}\times \hat{\mathbf{x}}].
\end{equation}
We note that $[\hat{\mathbf{K}}\times \hat{\mathbf{x}}]=\hat{\mathbf{y}}\sin\phi-\hat{\mathbf{z}}\cos\phi$.

It follows from Eqs.~\eqref{fl26a} and \eqref{fl26} that the ellipticity of the local electric polarization of the TM evanescent mode $\mu=(\omega\mathbf{K}p1)$ with $K>k$ arises as a consequence
of the fact that field in the TM evanescent mode has a longitudinal component that is aligned along the in-plane wave vector $\mathbf{K}$. The phase of this component
is shifted by $\pi/2$ from the phase of the transverse component that is aligned along the axis $x$.

Equation \eqref{fl27} shows that the local electric spin  density $\mathbf{S}_{\omega\mathbf{K}p1}$ is a vector that depends on the direction vector $\hat{\mathbf{K}}$ of the in-plane wave vector $\mathbf{K}$. In particular, a reverse of $\hat{\mathbf{K}}$ leads to a reverse of the electric spin density vector $\mathbf{S}_{\omega\mathbf{K}p1}$.
This is a signature of the so-called spin-orbit interaction of light \cite{Zeldovich,Bliokh review}. 
Thus, the difference between the rates of spontaneous emission into the evanescent modes with the opposite in-plane propagation directions 
$\mathbf{K}$ and $-\mathbf{K}$ is a consequence of spin-orbit coupling of light.

We observe from Eqs.~\eqref{fl25} and \eqref{fl27} that the local electric spin density $\mathbf{S}_{\omega_0\mathbf{K}p1}$ of the TM evanescent mode $\mu=(\omega_0\mathbf{K}p1)$ with $K>k_0$
and, consequently, the rate difference $\Delta F_{\mathrm{evan}}$ for evanescent modes with opposite in-plane propagation directions reduce exponentially with increasing
distance $x$ from the atom to the dielectric surface. For $x=0$, the magnitudes of $\mathbf{S}_{\omega_0\mathbf{K}p1}$ and $\Delta F_{\mathrm{evan}}$ achieve their maximum values,
which depend on $\xi$. In the limit $\xi\to0$, that is, $\kappa\to1$, we have $\mathbf{S}_{\omega_0\mathbf{K}p1}=0$ and, hence, $\Delta F_{\mathrm{evan}}=0$.

In order to get deep insight into the underlying physics of asymmetry between the rates of spontaneous emission into opposite in-plane propagation directions, we perform
the following general tensor analysis: It is clear that the rate $\gamma_{\alpha}$ of spontaneous emission into a mode $\alpha$ with the mode profile function $\mathbf{e}^{(\alpha)}$ is proportional
to the quantity $|\mathbf{d}_{eg}\cdot \mathbf{e}^{(\alpha)}|^2$, that is, 
\begin{equation}\label{fl28}
\gamma_{\alpha}=\mathcal{N}_{\alpha}|\mathbf{d}_{eg}\cdot \mathbf{e}^{(\alpha)}|^2,
\end{equation}
where $\mathcal{N}_{\alpha}$ is a parameter that does not depend on the relative orientation between $\mathbf{d}_{eg}$ and $\mathbf{e}^{(\alpha)}$. 
It follows from Eq.~\eqref{h11} of Appendix that we can decompose the rate $\gamma_{\alpha}$ as
\begin{equation}\label{fl29}
\gamma_{\alpha}=\gamma_{\alpha}^{(0)}+\gamma_{\alpha}^{(1)}+\gamma_{\alpha}^{(2)},
\end{equation}
where
\begin{subequations}\label{fl30}
\begin{align}
\gamma_{\alpha}^{(0)}&=\frac{\mathcal{N}_{\alpha}}{3}|\mathbf{d}_{eg}|^2|\mathbf{e}^{(\alpha)}|^2,\label{fl30a}\\
\gamma_{\alpha}^{(1)}&=\frac{\mathcal{N}_{\alpha}}{2}[\mathbf{d}_{eg}^*\times\mathbf{d}_{eg}]\cdot[\mathbf{e}^{(\alpha)*}\times\mathbf{e}^{(\alpha)}],\label{fl30b}\\
\gamma_{\alpha}^{(2)}&=\mathcal{N}_{\alpha}\{\mathbf{d}_{eg}^*\otimes\mathbf{d}_{eg}\}_2\cdot\{\mathbf{e}^{(\alpha)*}\otimes\mathbf{e}^{(\alpha)}\}_2.\label{fl30c}
\end{align}
\end{subequations}
In Eq.~\eqref{fl30c}, the notation $\{\mathbf{A}^*\otimes\mathbf{A}\}_{2}$ stands for the tensor product of rank 2 of the complex vectors $\mathbf{A}^*$ and $\mathbf{A}$.
The quantities $\gamma_{\alpha}^{(0)}$, $\gamma_{\alpha}^{(1)}$, and  $\gamma_{\alpha}^{(2)}$ are called the scalar, vector, and tensor components of the rate $\gamma_{\alpha}$, respectively.

According to Eq.~\eqref{fl30a}, the scalar component $\gamma_{\alpha}^{(0)}$ of the spontaneous emission rate does not depend on the orientations and circulations of the atomic dipole matrix element vector $\mathbf{d}_{eg}$ as well as the orientations and circulations of the field mode profile vector $\mathbf{e}^{(\alpha)}$. This component is the spontaneous emission rate averaged over the orientation of the dipole matrix element vector $\mathbf{d}_{eg}$ in space.

According to Eq.~\eqref{fl30b}, the vector component $\gamma_{\alpha}^{(1)}$ of the spontaneous emission rate depends on the overlap between the vectors $i[\mathbf{d}_{eg}^*\times\mathbf{d}_{eg}]$ and $-i[\mathbf{e}^{(\alpha)*}\times\mathbf{e}^{(\alpha)}]$, which are proportional to the ellipticity vector of the atomic electric dipole polarization and the ellipticity vector of the electric field polarization, respectively. 
The vector $i[\mathbf{d}_{eg}^*\times\mathbf{d}_{eg}]$ characterizes an effective magnetic dipole produced by the rotation of the electric dipole, and is responsible for the vector polarizability of the atom.
The vector $-i[\mathbf{e}^{(\alpha)*}\times\mathbf{e}^{(\alpha)}]$ characterizes an effective magnetic field and is responsible for the local electric spin  density of light. The vector component $\gamma_{\alpha}^{(1)}$ of the rate can be considered as a result of the interaction between the effective magnetic dipole and the effective magnetic field. Due to spin-orbit coupling of light \cite{Zeldovich,Bliokh review}, a reverse of the propagation direction leads to a reverse of the spin density of light and, consequently, to a reverse of the vector component $\gamma_{\alpha}^{(1)}$ of the spontaneous emission rate.

According to Eq.~\eqref{fl30c}, the tensor component $\gamma_{\alpha}^{(2)}$ of the spontaneous emission rate depends on the scalar product of the irreducible tensors $\{\mathbf{d}_{eg}^*\otimes\mathbf{d}_{eg}\}_2$ and $\{\mathbf{e}^{(\alpha)*}\otimes\mathbf{e}^{(\alpha)}\}_2$ for the atomic dipole and the field mode profile, respectively. The tensor $\{\mathbf{d}_{eg}^*\otimes\mathbf{d}_{eg}\}_2$ is responsible for the tensor polarizability of the atom. In general, $\{\mathbf{e}^{(\alpha)*}\otimes\mathbf{e}^{(\alpha)}\}_2$ and, hence $\gamma_{\alpha}^{(2)}$ depend on the azimuthal angle $\phi$ of the in-plane wave vector $\mathbf{K}$ in the $yz$ plane.
We can  show that, for the evanescent modes, in the half-space $x>0$, the tensor $\{\mathbf{e}^{(\alpha)*}\otimes\mathbf{e}^{(\alpha)}\}_2$ and, hence, the tensor component $\gamma_{\alpha}^{(2)}$ of the rate $\gamma_{\alpha}$ do not change when we reverse the direction of the in-plane wave vector $\mathbf{K}$.

We now calculate the rates of spontaneous emission into evanescent modes propagating into separate sides of a plane containing the axis $x$, on which the atom is located.
Without loss of generality, we choose the plane $xy$. 
The rates $\gamma_{\mathrm{evan}}^{(+)}$ and $\gamma_{\mathrm{evan}}^{(-)}$ of spontaneous emission into evanescent modes propagating into the $+z$ and $-z$ sides, respectively, are given by
\begin{equation}\label{fl31}
\begin{split}
\gamma_{\mathrm{evan}}^{(+)}&=\gamma_0\int\limits_{0}^{\sqrt{n_1^2-1}} \xi d\xi\int\limits_0^{\pi}F_{\mathrm{evan}}(\xi,\phi)\, d\phi,\\
\gamma_{\mathrm{evan}}^{(-)}&=\gamma_0\int\limits_{0}^{\sqrt{n_1^2-1}} \xi d\xi\int\limits_{\pi}^{2\pi}F_{\mathrm{evan}}(\xi,\phi)\, d\phi.
\end{split}
\end{equation}
We find
\begin{equation}\label{fl32}
\begin{split}
\gamma_{\mathrm{evan}}^{(+)}&=\frac{\gamma_{\mathrm{evan}}}{2}+\frac{\Delta_{\mathrm{evan}}}{2},\\
\gamma_{\mathrm{evan}}^{(-)}&=\frac{\gamma_{\mathrm{evan}}}{2}-\frac{\Delta_{\mathrm{evan}}}{2},
\end{split}
\end{equation}
where
\begin{equation}\label{fl33}
\begin{split}
\gamma_{\mathrm{evan}}&=\frac{3}{4}\gamma_0\int\limits_{0}^{\sqrt{n_1^2-1}}  \big\{(1-|u_x|^2)T_s(\xi)\\
&\quad + [|u_x|^2(2+\xi^2)+\xi^2]T_p(\xi) \big\}e^{-2\xi k_0x} d\xi
\end{split}
\end{equation}
is the rate of spontaneous emission into evanescent modes in all directions \cite{Agarwal,Wylie,Nano-Optics} and
\begin{equation}\label{fl34}
\Delta_{\mathrm{evan}}=\frac{6}{\pi}\gamma_0\,\mathrm{Im\,}(u_x^*u_z)\int\limits_{0}^{\sqrt{n_1^2-1}}
\xi\sqrt{1+\xi^2}\,T_p(\xi) e^{-2\xi k_0x} \,d\xi
\end{equation}
is the difference between the rate components $\gamma_{\mathrm{evan}}^{(+)}$ and $\gamma_{\mathrm{evan}}^{(-)}$ for the opposite sides $+z$ and $-z$, respectively.
It is clear from Eq.~\eqref{fl34} that the rate difference $\Delta_{\mathrm{evan}}$ depends on the imaginary part of the cross term $u_x^*u_z$, that is, on the ellipticity of the polarization of the atomic dipole vector in the $xz$ plane. Meanwhile, Eq. \eqref{fl33} shows that the rate $\gamma_{\mathrm{evan}}$ for all evanescent modes
does not depend on the ellipticity of the dipole polarization. 
We note that the sign (plus or minus) of the rate difference $\Delta_{\mathrm{evan}}$ for evanescent modes is determined by the sign of $\mathrm{Im\,}(u_x^*u_z)$ and, hence,
does not depend on the distance $x$. In the limit $x\to\infty$, we have $\gamma_{\mathrm{evan}}\to0$ and $\Delta_{\mathrm{evan}}\to0$.
When the dipole polarization vector $\mathbf{u}$ is a real vector, the rate difference for evanescent modes propagating into the opposite sides $+z$ and $-z$ is $\Delta_{\mathrm{evan}}=0$. 

The asymmetry between the rates $\gamma_{\mathrm{evan}}^{(+)}$ and $\gamma_{\mathrm{evan}}^{(-)}$ for  the $+z$ and $-z$ sides, respectively, is characterized by the factor 
$\zeta_{\mathrm{evan}}=\Delta_{\mathrm{evan}}/\gamma_{\mathrm{evan}}$. It is interesting to note that, unlike the asymmetry factor
$\zeta_{F_\mathrm{evan}}$ for the angular rate densities $F_{\mathrm{evan}}(\xi,\phi)$ and $F_{\mathrm{evan}}(\xi,\phi+\pi)$, the asymmetry factor $\zeta_{\mathrm{evan}}$ for the side rates $\gamma_{\mathrm{evan}}^{(+)}$ and $\gamma_{\mathrm{evan}}^{(-)}$ depends on the distance $x$. The reason is that, according to Eqs.~\eqref{fl34} and \eqref{fl33}, the difference $\Delta_{\mathrm{evan}}$ between and the sum $\gamma_{\mathrm{evan}}$ of the side rates
$\gamma_{\mathrm{evan}}^{(+)}$ and $\gamma_{\mathrm{evan}}^{(-)}$ are given by different integrals over the variable $\xi$. 
The kernels of these integrals are different from each other although they contain a common exponential factor $e^{-2\xi k_0x}$.
Due to the integration over $\xi$, the $x$ dependence of $\Delta_{\mathrm{evan}}$ is different from that of $\gamma_{\mathrm{evan}}$. Consequently, the asymmetry factor $\zeta_{\mathrm{evan}}=\Delta_{\mathrm{evan}}/\gamma_{\mathrm{evan}}$ for the side rates $\gamma_{\mathrm{evan}}^{(+)}$ and $\gamma_{\mathrm{evan}}^{(-)}$ is a function of the distance $x$. 

In the particular case where the dipole matrix element vector $\mathbf{d}_{eg}$ is perpendicular to the interface, we obtain \cite{Agarwal,Wylie,Nano-Optics}
\begin{equation}
\gamma_{\mathrm{evan}}=\gamma_{\mathrm{evan}}^{\perp}=
\frac{3}{2}\gamma_0\int\limits_0^{\sqrt{n_1^2-1}}T_{\perp}(\xi)e^{-2\xi k_0 x}d\xi,
\label{fl35}
\end{equation}
and, in the particular case where the dipole matrix element vector $\mathbf{d}_{eg}$ lies in the interface plane $yz$, we find \cite{Agarwal,Wylie,Nano-Optics}
\begin{equation}
\gamma_{\mathrm{evan}}=\gamma_{\mathrm{evan}}^{\parallel}=
\frac{3}{4}\gamma_0\int\limits_0^{\sqrt{n_1^2-1}}T_{\parallel}(\xi)e^{-2\xi k_0 x}d\xi.
\label{fl36}
\end{equation}
Here, we have introduced the parameters $T_{\perp}=(1+\xi^2)T_p$ and $T_{\parallel}=T_s+\xi^2 T_p$, whose explicit expressions are
\begin{eqnarray}
T_{\perp}&=&\frac{2n_1^2}{n_1^2-1}\,
\frac{\sqrt{n_1^2-1-\xi^2}}{(n_1^2+1)\xi^2+1}\,\xi(1+\xi^2),
\nonumber\\ 
T_{\parallel}&=&
\frac{2}{n_1^2-1}
\left(1+\frac{n_1^2\xi^2}{(n_1^2+1)\xi^2+1}\right)
\xi\sqrt{n_1^2-1-\xi^2}.\qquad
\label{fl37}
\end{eqnarray}
In both cases, we have $\Delta_{\mathrm{evan}}=0$. 

\subsection{Spontaneous emission into radiation modes}
\label{subsec:radiation}

The rate of spontaneous emission from the atom at a position $x>0$ into radiation modes is
\begin{equation}\label{fl38}
\gamma_{\mathrm{rad}}=\sum_{q=s,p}\sum_{j=1,2}\gamma_{\mathrm{rad}}^{qj},
\end{equation}
where the notation
\begin{equation}\label{fl39}
\gamma_{\mathrm{rad}}^{qj}=2\pi\int\limits_{0}^{k_0}K\, dK\int\limits_0^{2\pi}|G_{\omega_0\mathbf{K}qj}|^2\, d\phi
\end{equation}
with $q=s,p$ and $j=1,2$ stands for the rate of spontaneous emission into the $qj$-type radiation modes.

We again use the notation $\kappa=K/k_0$ for the normalized magnitude of the in-plane component $\mathbf{K}$ of the wave vector 
and the notation $\xi=\sqrt{|1-\kappa^2|}$ for the normalized magnitude of the out-of-plane component $\beta_2\hat{\mathbf{x}}$ of the wave vector in the half-space $x>0$.
For radiation modes, we have $\beta_2=k_0\xi$, $0\le \kappa\le 1$, and $0\le \xi=\sqrt{1-\kappa^2}\le 1$.  
We change the integration variable of the first integral in Eq.~\eqref{fl39} from $K$ to $\xi$.
Then, we obtain
\begin{equation}\label{fl40}
\begin{split}
\gamma_{\mathrm{rad}}&=\gamma_0\int\limits_{0}^{1}\xi d\xi\int\limits_0^{2\pi}F_{\mathrm{rad}}(\xi,\phi)\, d\phi,\\
\gamma_{\mathrm{rad}}^{qj}&=\gamma_0\int\limits_{0}^{1}\xi d\xi\int\limits_0^{2\pi}F_{\mathrm{rad}}^{qj}(\xi,\phi)\, d\phi,
\end{split}
\end{equation}
where 
\begin{equation}\label{fl40a}
F_{\mathrm{rad}}=\sum_{q=s,p}\sum_{j=1,2}F_{\mathrm{rad}}^{qj}, 
\end{equation}
with
\begin{eqnarray}\label{f12a}
F_{\mathrm{rad}}^{s1}&=&\frac{3}{8\pi\xi} (1- r_s^2) \big[|u_y|^2\sin^2\phi+|u_z|^2\cos^2\phi \nonumber\\
&&\mbox{} -\mathrm{Re\,}(u_y^*u_z)\sin2\phi\big],       
\end{eqnarray}
\begin{eqnarray}\label{f12b}
F_{\mathrm{rad}}^{s2}&=&\frac{3}{8\pi\xi}\big[1+r_s^2+2r_s\cos(2\xi k_0x)\big] \big[|u_y|^2\sin^2\phi \nonumber\\
&&\mbox{}+|u_z|^2\cos^2\phi -\mathrm{Re\,}(u_y^*u_z)\sin2\phi\big],       
\end{eqnarray}
\begin{eqnarray}\label{f14a}
F_{\mathrm{rad}}^{p1}&=&\frac{3}{8\pi\xi}\big\{(1-r_p^2)[|u_x|^2(1-\xi^2)+|u_y|^2\xi^2\cos^2\phi \nonumber\\
&&\mbox{}  +|u_z|^2\xi^2\sin^2\phi + \mathrm{Re\,}(u_y^*u_z)\xi^2\sin2\phi] \nonumber\\
&&\mbox{}  - 2(1-r_p^2)\xi\sqrt{1-\xi^2}\,\mathrm{Re\,}(u_x^*u_y\cos\phi + u_x^*u_z\sin\phi)\big\},\nonumber\\
\end{eqnarray}
and
\begin{eqnarray}\label{f14b}
F_{\mathrm{rad}}^{p2}&=&\frac{3}{8\pi\xi}\big\{(1+r_p^2)[|u_x|^2(1-\xi^2)+|u_y|^2\xi^2\cos^2\phi \nonumber\\
&&\mbox{}  +|u_z|^2\xi^2\sin^2\phi + \mathrm{Re\,}(u_y^*u_z)\xi^2\sin2\phi] \nonumber\\
&&\mbox{}  + 2(1-r_p^2)\xi\sqrt{1-\xi^2}\,\mathrm{Re\,}(u_x^*u_y\cos\phi + u_x^*u_z\sin\phi)\big\}\nonumber\\
&&\mbox{}  + \frac{3}{4\pi\xi}r_p \big\{\cos(2\xi k_0x)\big[|u_x|^2(1-\xi^2)-|u_y|^2\xi^2\cos^2\phi \nonumber\\
&&\mbox{} - |u_z|^2\xi^2\sin^2\phi  - \mathrm{Re\,}(u_y^*u_z)\xi^2\sin2\phi\big] \nonumber\\
&&\mbox{} + 2\xi\sqrt{1-\xi^2}\sin(2\xi k_0x)\nonumber\\
&&\mbox{} \times \mathrm{Im\,}(u_x^*u_y\cos\phi + u_x^*u_z\sin\phi) \big\}.
\end{eqnarray}
Here, we have introduced the notations $r_s\equiv r_{21}^s=(\xi-\eta)/(\xi+\eta)$ 
and $r_p\equiv r_{21}^p=(n_1^2\xi-\eta)/(n_1^2\xi+\eta)$ for the reflection coefficients of light coming from medium 2 to medium 1,  
where $\eta\equiv\sqrt{n_1^2-\kappa^2}=\sqrt{n_1^2-1+\xi^2}$.
The explicit expressions for the reflection coefficients $r_s$ and $r_p$ are given in terms of $\xi$ as
\begin{equation}\label{fl43}
\begin{split}
r_s&=\frac{\xi-\sqrt{n_1^2-1+\xi^2}}{\xi+\sqrt{n_1^2-1+\xi^2}},\\
r_p&=\frac{n_1^2\xi-\sqrt{n_1^2-1+\xi^2}}{n_1^2\xi+\sqrt{n_1^2-1+\xi^2}}.    
\end{split}
\end{equation}

In the half-space $x>0$, the wave vector of a radiation mode is $(\beta_2,K_y,K_z)$, where $\beta_2=k_0\xi$. The parameters $\xi$ and $\kappa=\sqrt{1-\xi^2}$ and the angle $\phi$ characterize the components of the wave vector $(\beta_2,K_y,K_z)$ of a radiation mode in the half-space $x>0$ via the relations $\beta_2/k_0=\xi$, $K_y/k_0=\kappa_y=\kappa\cos\phi$, and $K_z/k_0=\kappa_z=\kappa\sin\phi$. 

The functions $F_{\mathrm{rad}}^{sj}$ and $F_{\mathrm{rad}}^{pj}$ are respectively the angular densities of the spontaneous emission rates into the radiation modes $\nu=(\omega_0\mathbf{K}sj)$ and $(\omega_0\mathbf{K}pj)$, with $0\le K\leq k_0$, in the wave vector space. The function $F_{\mathrm{rad}}$ is the angular density of the spontaneous emission rate into both $s$ and $p$ types of radiation modes. 
In the limit $\kappa\to1$, that is, $K\to k_0$, we have
\begin{eqnarray}\label{fl44}
\lim_{\kappa\to1}F_{\mathrm{rad}}&=&\frac{3}{2\pi}\frac{1}{\sqrt{n_1^2-1}} [n_{1}^2|u_x|^2+|u_y|^2\sin^2\phi\nonumber\\
&&\mbox{}+|u_z|^2\cos^2\phi -\mathrm{Re\,}(u_y^*u_z)\sin2\phi].      
\end{eqnarray}
Comparison between Eqs.~\eqref{fl44} and \eqref{fl21} confirms that $\lim_{\kappa\to1}F_{\mathrm{rad}}=\lim_{\kappa\to1}F_{\mathrm{evan}}$.

We introduce the notations $F_{\mathrm{rad}}^{s}=F_{\mathrm{rad}}^{s1}+F_{\mathrm{rad}}^{s2}$ and $F_{\mathrm{rad}}^{p}=F_{\mathrm{rad}}^{p1}+F_{\mathrm{rad}}^{p2}$, which are the angular densities of the spontaneous emission rates into the radiation modes of the $s$ and $p$ types, respectively. We find
\begin{eqnarray}\label{fl41}
F_{\mathrm{rad}}^{s}&=&\frac{3}{4\pi\xi}\big[1+r_s\cos(2\xi k_0x)\big] \big[|u_y|^2\sin^2\phi+|u_z|^2\cos^2\phi \nonumber\\
&&\mbox{} -\mathrm{Re\,}(u_y^*u_z)\sin2\phi\big]       
\end{eqnarray}
and 
\begin{eqnarray}\label{fl42}
\lefteqn{F_{\mathrm{rad}}^p=\frac{3}{4\pi\xi}\big[|u_x|^2(1-\xi^2)+|u_y|^2\xi^2\cos^2\phi }\nonumber\\
&&\mbox{} + |u_z|^2\xi^2\sin^2\phi + \mathrm{Re\,}(u_y^*u_z)\xi^2\sin2\phi\big] \nonumber\\
&&\mbox{} + \frac{3}{4\pi\xi}r_p \big\{\cos(2\xi k_0x)\big[|u_x|^2(1-\xi^2)-|u_y|^2\xi^2\cos^2\phi \nonumber\\
&&\mbox{} - |u_z|^2\xi^2\sin^2\phi  - \mathrm{Re\,}(u_y^*u_z)\xi^2\sin2\phi\big] \nonumber\\
&&\mbox{} + 2\xi\sqrt{1-\xi^2}\sin(2\xi k_0x)\,\mathrm{Im\,}(u_x^*u_y\cos\phi + u_x^*u_z\sin\phi) \big\}.\nonumber\\
\end{eqnarray}
It is clear that $F_{\mathrm{rad}}=F_{\mathrm{rad}}^{s}+F_{\mathrm{rad}}^{p}$.

The mode function $\mathbf{U}_{\alpha}$, given by Eqs.~\eqref{fl1} and \eqref{fl2}, describes the mode $\alpha=(\omega\mathbf{K}qj)$, which has a single input incident from medium $j$ to the interface. 
The function $\mathbf{U}_{\tilde{\alpha}}=\mathbf{U}_{\alpha}^*$ describes the mode 
$\tilde{\alpha}=(\omega,-\mathbf{K},q,\tilde{j})$, which has a single output coming from the interface into medium $j$.
The density $F_{\mathrm{rad}}^{q\tilde{j}}$ of the rate of spontaneous emission into a single-output mode $(\omega\mathbf{K}q\tilde{j})$ can be obtained from that for the single-input mode $(\omega,-\mathbf{K},q,j)$ by replacing the dipole polarization vector $\mathbf{u}$ with its complex conjugate vector $\mathbf{u}^*$, that is, by applying the transformation $\mathcal{T}=(\mathbf{u}\to\mathbf{u}^*,\phi\to\phi+\pi)$ to $F_{\mathrm{rad}}^{qj}$. The transformation $\mathcal{T}$ does not change the rate density functions $F_{\mathrm{rad}}^{sj}$ [see Eqs.~\eqref{f12a} and \eqref{f12b}], $F_{\mathrm{rad}}^{s}$ [see Eq.~\eqref{fl41}],  $F_{\mathrm{rad}}^{p}$ [see Eq.~\eqref{fl42}], and $F_{\mathrm{rad}}$.
However, the transformation $\mathcal{T}$ reverses the sign of the term in the last line of Eq.~\eqref{f14a} and the term in the
third line of Eq.~\eqref{f14b} for $F_{\mathrm{rad}}^{p1}$ and $F_{\mathrm{rad}}^{p2}$, respectively. 
These terms cancel each other and therefore do not appear in the expressions for $F_{\mathrm{rad}}^{p}$ and $F_{\mathrm{rad}}$.
Thus, the functions $F_{\mathrm{rad}}^{sj}$, $F_{\mathrm{rad}}^{s}$,  $F_{\mathrm{rad}}^{p}$, and $F_{\mathrm{rad}}$ describe the distributions of the emission rates not only for single-input modes but also for singe-output modes.  

We observe that all the terms in expression \eqref{fl41} are associated with the coefficients $\sin^2\phi$, $\cos^2\phi$, and $\sin2\phi$, which do not vary with respect to the transformation $\phi\to\phi+\pi$. Hence, the rate density $F_{\mathrm{rad}}^{s}$ for the TE radiation modes has the same value for the opposite in-plane propagation directions $\mathbf{K}$ and $-\mathbf{K}$. 
Meanwhile, the terms in the last line of expression \eqref{fl42} contain the coefficients
$\cos\phi$ and $\sin\phi$,  which change their sign when we replace $\phi$ by $\phi+\pi$.
Hence, the rate density $F_{\mathrm{rad}}^{p}$ for the TM radiation modes may take different values for the opposite in-plane  propagation directions $\mathbf{K}$ and $-\mathbf{K}$. 
This asymmetry in spontaneous emission occurs when either $\mathrm{Im\,}(u_x^*u_y)$ or $\mathrm{Im\,}(u_x^*u_z)$ is not zero, 
that is, when the atomic dipole polarization vector $\mathbf{u}$  is a complex vector in a plane containing the axis $x$. 
As already mentioned, the fact that $\mathbf{u}$ is a complex vector means that the direction of the dipole of the atom rotates with time in space. The asymmetry of spontaneous emission into radiation modes with respect to central inversion in the interface plane appears as a consequence of the interference between  
the emission from the out-of-plane dipole component $u_x$ and the emission from the in-plane dipole components $u_y$ and $u_z$ where $u_x$ has a phase lag with respect to $u_y$ or $u_z$.
When the dipole polarization vector $\mathbf{u}$ is a real vector, the rate density $F_{\mathrm{rad}}$ for radiation modes is symmetric with respect to central inversion in the interface plane.
We note that, according to Eq.~\eqref{fl44}, in the limit $\kappa\to1$, the rate density $F_{\mathrm{rad}}$ is symmetric with respect to central inversion in the interface plane
for an arbitrary dipole polarization vector $\mathbf{u}$.

It is clear from Eqs.~\eqref{fl41} and \eqref{fl42} that the difference $\Delta F_{\mathrm{rad}}\equiv F_{\mathrm{rad}}(\xi,\phi)-F_{\mathrm{rad}}(\xi,\phi+\pi)$ between the rate densities $F_{\mathrm{rad}}(\xi,\phi)$ and $F_{\mathrm{rad}}(\xi,\phi+\pi)$ of spontaneous emission into the radiation modes with the opposite in-plane wave vectors $\mathbf{K}$ and $-\mathbf{K}$ is 
\begin{equation}\label{fl45}
\begin{split}
\Delta F_{\mathrm{rad}}&=
\frac{3}{\pi}\sqrt{1-\xi^2}\,r_p\sin(2\xi k_0x)\\
&\quad\times\mathrm{Im\,}(u_x^*u_y\cos\phi + u_x^*u_z\sin\phi).
\end{split}
\end{equation}
We note that the sign (plus or minus) of the rate  density difference $\Delta F_{\mathrm{rad}}$ for radiation modes depends on not only 
the dipole polarization vector $\mathbf{u}$ and the emission azimuthal angle $\phi$ but also on the  atom-interface distance $x$ 
and the out-of-plane wave-vector-component  parameter $\xi$. 
When the dipole polarization vector $\mathbf{u}$ is a real vector, the rate density difference for radiation modes with opposite in-plane wave vectors is $\Delta F_{\mathrm{rad}}=0$.

The asymmetry degree of the angular density $F_{\mathrm{rad}}$ under central inversion in the interface plane is characterized by the factor 
$\zeta_{F_\mathrm{rad}}=\Delta F_{\mathrm{rad}}/F_{\mathrm{rad}}^{\mathrm{sum}}$, where 
$F_{\mathrm{rad}}^{\mathrm{sum}}\equiv F_{\mathrm{rad}}(\xi,\phi)+F_{\mathrm{rad}}(\xi,\phi+\pi)$. It is clear that the asymmetry factor
$\zeta_{F_\mathrm{rad}}$ depends on not only $\xi$ and $\phi$ but also $x$.

We can easily show that
\begin{equation}\label{fl46}
\Delta F_{\mathrm{rad}}=\frac{3}{8\pi\xi}[\mathbf{u}^*\times\mathbf{u}]\cdot[\mathbf{U}_{\omega_0\mathbf{K}p2}^{*}\times\mathbf{U}_{\omega_0\mathbf{K}p2}].
\end{equation}
As already mentioned, the vector $i[\mathbf{u}^*\times\mathbf{u}]$ is the ellipticity vector of the atomic dipole polarization. Meanwhile, the vector $-i[\mathbf{U}_{\omega_0\mathbf{K}p2}^*\times\mathbf{U}_{\omega_0\mathbf{K}p2}]$ is proportional to the ellipticity vector of the electric polarization and, consequently, to the electric spin density vector of the TM radiation mode $\nu=(\omega_0\mathbf{K}p2)$ at the position of the atom. Equation \eqref{fl46} indicates that the difference $\Delta F_{\mathrm{rad}}$ is a result of the overlap between the ellipticity vector of the atomic dipole polarization and the ellipticity vector of the local electric polarization of the TM radiation mode $\nu=(\omega_0\mathbf{K}p2)$. The electric parts of the other radiation modes, that is, the modes $\nu=(\omega_0\mathbf{K}s1)$, $(\omega_0\mathbf{K}s2)$, and $(\omega_0\mathbf{K}p1)$, with $K\le k_0$, are linearly polarized in the half-space $x>0$. These modes do not contribute to $\Delta F_{\mathrm{rad}}$.

Comparison between Eqs.~\eqref{fl46} and \eqref{fl24} shows that 
\begin{equation}\label{fl47}
\Delta F_{\mathrm{rad}}=\frac{3\omega_0}{2\pi\epsilon_0\xi} i[\mathbf{u}^*\times\mathbf{u}]\cdot\mathbf{S}_{\omega_0\mathbf{K}p2},
\end{equation}
where $\mathbf{S}_{\omega_0\mathbf{K}p2}$ is the local electric spin  density of the field in the TM radiation mode $\nu=(\omega_0\mathbf{K}p2)$ with the positive-frequency-component envelope 
$\boldsymbol{\mathcal{E}}_{\omega_0\mathbf{K}p2}=\mathbf{U}_{\omega_0\mathbf{K}p2}$. 
In the half-space $x>0$, where the atom is located, the electric polarization vector of
the TM radiation mode $\nu=(\omega\mathbf{K}p2)$ is 
\begin{equation}\label{fl48}
\boldsymbol{\epsilon}_{\omega\mathbf{K}p2}=\frac{1}{\sqrt{Z}}
[\kappa\hat{\mathbf{x}}+\xi\hat{\mathbf{K}}+r_p e^{2i\xi kx} (\kappa\hat{\mathbf{x}}-\xi\hat{\mathbf{K}})],
\end{equation}
where $Z=1+r_p^2+2r_p(1-2\xi^2)\cos(2\xi kx)$. The ellipticity vector of the electric polarization of the mode is found to be
\begin{equation}\label{fl49}
\mathrm{Im}\big[\boldsymbol{\epsilon}_{\omega\mathbf{K}p2}^{*}\times\boldsymbol{\epsilon}_{\omega\mathbf{K}p2}\big]
=\frac{4}{Z}\xi\sqrt{1-\xi^2}\,r_p\sin(2\xi kx)[\hat{\mathbf{K}}\times \hat{\mathbf{x}}],
\end{equation}
which leads to the local electric spin density
\begin{equation}\label{fl50}
\begin{split}
\mathbf{S}_{\omega\mathbf{K}p2}&=\frac{\epsilon_0}{\omega}\xi\sqrt{1-\xi^2}\frac{n_1^2\xi-\sqrt{n_1^2-1+\xi^2}}{n_1^2\xi+\sqrt{n_1^2-1+\xi^2}}\sin(2\xi kx)\\
&\quad \times[\hat{\mathbf{K}}\times \hat{\mathbf{x}}].
\end{split}
\end{equation}
We note that $[\hat{\mathbf{K}}\times \hat{\mathbf{x}}]=\hat{\mathbf{y}}\sin\phi-\hat{\mathbf{z}}\cos\phi$.

It follows from Eqs.~\eqref{fl48} and \eqref{fl49} that the ellipticity of the local electric polarization of the TM radiation mode $\nu=(\omega\mathbf{K}p2)$ arises as a consequence
of the change of the polarization vector from $\kappa\hat{\mathbf{x}}+\xi\hat{\mathbf{K}}$ to $\kappa\hat{\mathbf{x}}-\xi\hat{\mathbf{K}}$ due to the reflection, the additional phase $2\xi kx$ of the reflected beam due to a round trip between the point $x$ and the interface, and the interference between the incident and reflected beams.
We note that the reflection leads to a change of the electric polarization vector in the case where the electric component of the field lies in the incidence plane, that is, the case
of $p$ modes.

Equation \eqref{fl50} shows that a reverse of $\hat{\mathbf{K}}$ leads to a reverse of the spin density vector $\mathbf{S}_{\omega\mathbf{K}p2}$.
This is a signature of spin-orbit coupling of light \cite{Zeldovich,Bliokh review}. 
The difference between the rates of spontaneous emission into the radiation modes with the opposite in-plane propagation directions 
$\mathbf{K}$ and $-\mathbf{K}$ is a consequence of spin-orbit coupling of light \cite{Zeldovich,Bliokh review}, like in the case of evanescent modes.

We observe from Eqs.~\eqref{fl50} and \eqref{fl47} that the local electric spin density $\mathbf{S}_{\omega_0\mathbf{K}p2}$ of the TM radiation mode $\nu=(\omega_0\mathbf{K}p2)$ and, consequently, the rate difference $\Delta F_{\mathrm{rad}}$ for radiation modes with opposite in-plane propagation directions oscillate as $\sin(2\xi k_0x)$ with increasing
distance $x$ from the atom to the dielectric surface. For $x=0$, we have $\mathbf{S}_{\omega_0\mathbf{K}p2}=0$ and, hence, $\Delta F_{\mathrm{rad}}=0$.
This result is in contrast to the result for the case of evanescent modes, where the magnitudes of the spin density $\mathbf{S}_{\omega_0\mathbf{K}p1}$ for the TM evanescent mode $\mu=(\omega_0\mathbf{K}p1)$ with $K>k_0$ and, hence, the rate difference $\Delta F_{\mathrm{evan}}$ achieve their maximum values at the interface.
The explanation for the fact that $\Delta F_{\mathrm{rad}}=0$ at $x=0$ is simple. Indeed, at the interface, the relative phase between the incident light and the reflected light is 
just the phase of the reflection coefficient $r_p$. This phase is equal to 0 or $\pi$ when the incidence angle $\theta=\arccos(\xi)$ is smaller or larger
than the Brewster angle $\theta_B=\arctan(n_1)$, respectively.
Due to this fact, the ellipticity of the local electric polarization of the TM radiation mode $\nu=(\omega_0\mathbf{K}p2)$ and, hence, the rate density difference $\Delta F_{\mathrm{rad}}$
vanish at $x=0$. 

We now calculate the rates of spontaneous emission into radiation modes propagating into separate sides of a plane containing the axis $x$.
To be specific, we choose again the plane $xy$, as in the case of evanescent modes. 
The rates $\gamma_{\mathrm{rad}}^{(+)}$ and $\gamma_{\mathrm{rad}}^{(-)}$ of spontaneous emission into radiation modes propagating into the $+z$ and $-z$ sides, respectively, are given by
\begin{equation}\label{fl52}
\begin{split}
\gamma_{\mathrm{rad}}^{(+)}&=\gamma_0\int\limits_{0}^{1}\xi d\xi\int\limits_0^{\pi}F_{\mathrm{rad}}(\xi,\phi)\, d\phi,\\
\gamma_{\mathrm{rad}}^{(-)}&=\gamma_0\int\limits_{0}^{1}\xi d\xi\int\limits_{\pi}^{2\pi}F_{\mathrm{rad}}(\xi,\phi)\, d\phi.
\end{split}
\end{equation}
We can show that
\begin{equation}\label{fl53}
\begin{split}
\gamma_{\mathrm{rad}}^{(+)}&=\frac{\gamma_{\mathrm{rad}}}{2}+\frac{\Delta_{\mathrm{rad}}}{2},\\
\gamma_{\mathrm{rad}}^{(-)}&=\frac{\gamma_{\mathrm{rad}}}{2}-\frac{\Delta_{\mathrm{rad}}}{2},
\end{split}
\end{equation}
where
\begin{eqnarray}\label{fl54}
\gamma_{\mathrm{rad}}&=&\gamma_0
+\frac{3}{4}\gamma_0\int\limits_{0}^{1}\big\{ (1-|u_x|^2)r_s(\xi)  \nonumber\\
&&\mbox{} + \big[|u_x|^2(2-\xi^2)-\xi^2\big]r_p(\xi)\big\}\cos(2\xi k_0x) \,d\xi\qquad
\end{eqnarray}
is the rate of spontaneous emission into radiation modes in all directions \cite{Agarwal,Wylie,Nano-Optics} and
\begin{equation}\label{fl55}
\Delta_{\mathrm{rad}}=\frac{6}{\pi}\gamma_0\,\mathrm{Im\,}(u_x^*u_z)\int\limits_{0}^{1}\xi\sqrt{1-\xi^2}\,r_p(\xi)\sin(2\xi k_0x)\,d\xi
\end{equation}
is the difference between the rate components $\gamma_{\mathrm{rad}}^{(+)}$ and $\gamma_{\mathrm{rad}}^{(-)}$ for the opposite sides $+z$ and $-z$, respectively.
It is clear from Eq.~\eqref{fl55} that, like the rate difference $\Delta_{\mathrm{evan}}$ for evanescent modes, the rate difference $\Delta_{\mathrm{rad}}$ for radiation modes depends on the imaginary part of the cross term $u_x^*u_z$, that is, on the ellipticity of the polarization of the atomic dipole vector in the $xz$ plane. 
Meanwhile, Eq.~\eqref{fl54} shows that, like the rate $\gamma_{\mathrm{evan}}$ for evanescent modes, the rate $\gamma_{\mathrm{rad}}$ for radiation modes
does not depend on the ellipticity of the dipole polarization. We note that the sign (plus or minus) of the rate difference $\Delta_{\mathrm{rad}}$ for radiation modes depends on the distance $x$.
In the limit $x\to\infty$, we have $\gamma_{\mathrm{rad}}\to\gamma_0$ and $\Delta_{\mathrm{rad}}\to0$.
When the dipole polarization vector $\mathbf{u}$ is a real vector, the rate difference for radiation modes propagating into the opposite sides $+z$ and $-z$ is $\Delta_{\mathrm{rad}}=0$. 

The asymmetry between the rates $\gamma_{\mathrm{rad}}^{(+)}$ and $\gamma_{\mathrm{rad}}^{(-)}$ for  the $+z$ and $-z$ sides, respectively, is characterized by the factor 
$\zeta_{\mathrm{rad}}=\Delta_{\mathrm{rad}}/\gamma_{\mathrm{rad}}$. It is clear that the asymmetry factor $\zeta_{\mathrm{rad}}$ for the side rates $\gamma_{\mathrm{rad}}^{(+)}$ and $\gamma_{\mathrm{rad}}^{(-)}$ reduces to zero in the limit $x\to\infty$.

In the particular case where the dipole matrix element vector $\mathbf{d}_{eg}$ is perpendicular to the interface, we obtain \cite{Agarwal,Wylie,Nano-Optics}
\begin{equation}
\gamma_{\mathrm{rad}}=\gamma_{\mathrm{rad}}^{\perp}=
\gamma_0+\frac{3}{2}\gamma_0\int\limits_0^1 r_{\perp}(\xi)\cos(2\xi k_0 x)d\xi
\label{fl56}
\end{equation}
and, in the particular case where the dipole matrix element vector $\mathbf{d}_{eg}$ lies in the interface plane $yz$, we find \cite{Agarwal,Wylie,Nano-Optics}
\begin{equation}
\gamma_{\mathrm{rad}}=\gamma_{\mathrm{rad}}^{\parallel}=\gamma_0+\frac{3}{4}\gamma_0\int\limits_0^1 r_{\parallel}(\xi)\cos(2\xi k_0 x)d\xi.
\label{fl57}
\end{equation}
Here, we have introduced the parameters $r_{\perp}=(1-\xi^2) r_p$ and $r_{\parallel}=r_s-\xi^2 r_p$, whose explicit expressions are
\begin{eqnarray}
r_{\perp}&=&(1-\xi^2)\frac{n_{1}^2\xi-\sqrt{n_1^2-1+\xi^2}}
{n_{1}^2\xi+\sqrt{n_1^2-1+\xi^2}},
\nonumber\\
r_{\parallel}&=&
\frac{\xi-\sqrt{n_1^2-1+\xi^2}}{\xi+\sqrt{n_1^2-1+\xi^2}}
-\xi^2\frac{n_1^2\xi-\sqrt{n_1^2-1+\xi^2}}{n_1^2\xi+\sqrt{n_1^2-1+\xi^2}}.\qquad
\label{fl58}
\end{eqnarray}
In both cases, we have $\Delta_{\mathrm{rad}}=0$. 
The terms that contain the integrals in Eqs. (\ref{fl56}) and (\ref{fl57}) 
are the results of the interference between the emitted and reflected fields.

In order to derive the rates $\gamma^{(+)}=\gamma_{\mathrm{evan}}^{(+)}+\gamma_{\mathrm{rad}}^{(+)}$ and $\gamma^{(-)}=\gamma_{\mathrm{evan}}^{(-)}+\gamma_{\mathrm{rad}}^{(-)}$ of spontaneous emission into both evanescent and radiation types of modes propagating into the $+z$ and $-z$ sides, respectively, we sum up Eqs.~\eqref{fl32} and \eqref{fl53}. Then, we obtain
\begin{equation}\label{fl59}
\begin{split}
\gamma^{(+)}&=\frac{\gamma}{2}+\frac{\Delta}{2},\\
\gamma^{(-)}&=\frac{\gamma}{2}-\frac{\Delta}{2},
\end{split}
\end{equation}
where
\begin{eqnarray}\label{fl60}
\gamma&=&\gamma_0
+\frac{3}{4}\gamma_0\int\limits_{0}^{\sqrt{n_1^2-1}}  \big\{(1-|u_x|^2)T_s(\xi)\nonumber\\
&&\mbox{}+[|u_x|^2(2+\xi^2)+\xi^2]T_p(\xi) \big\}e^{-2\xi k_0x} d\xi \nonumber\\
&&\mbox{}
+\frac{3}{4}\gamma_0\int\limits_{0}^{1}\big\{ (1-|u_x|^2)r_s(\xi)  \nonumber\\
&&\mbox{} + \big[|u_x|^2(2-\xi^2)-\xi^2\big]r_p(\xi)\big\}\cos(2\xi k_0x) \,d\xi \qquad
\end{eqnarray}
is the total rate of spontaneous emission \cite{Agarwal,Wylie,Nano-Optics} and
\begin{eqnarray}\label{fl61}
\Delta&=&\frac{6}{\pi}\gamma_0\,\mathrm{Im\,}(u_x^*u_z)\Bigg[\int\limits_{0}^{\sqrt{n_1^2-1}}
\xi\sqrt{1+\xi^2}\,T_p(\xi) e^{-2\xi k_0x} \,d\xi \nonumber\\
&&\mbox{}+\int\limits_{0}^{1}
\xi\sqrt{1-\xi^2}\,r_p(\xi)\sin(2\xi k_0x)\,d\xi\Bigg]
\end{eqnarray}
is the difference between the rate components $\gamma^{(+)}$ and $\gamma^{(-)}$ for the opposite sides $+z$ and $-z$, respectively.
When the dipole polarization vector $\mathbf{u}$ is a real vector, the rate difference for both evanescent and radiation types of modes propagating into the opposite sides $+z$ and $-z$ is $\Delta=0$. 
The asymmetry between the rates $\gamma^{(+)}$ and $\gamma^{(-)}$ of directional spontaneous emission into both types of modes is characterized by the parameter $\zeta=\Delta/\gamma$.

\subsection{Spontaneous emission into radiation modes with outputs on a given side of the interface}
\label{subsec:dielectric and vacuum}

The function $F_{\mathrm{evan}}$, calculated in Sec.~\ref{subsec:evanescent}, is the density of the rate of spontaneous emission into evanescent modes, which have outputs in the dielectric. The function $F_{\mathrm{rad}}$, calculated in Sec.~\ref{subsec:radiation}, is the density of the rate of spontaneous emission into radiation modes with outputs on both sides of the interface. In this subsection, we consider the densities of the rates of spontaneous emission into radiation modes with outputs on a given side of the interface.

Let $F_{\mathrm{rad}}^{\mathrm{mat}}$ and $F_{\mathrm{rad}}^{\mathrm{vac}}$
be the angular densities of the rates of spontaneous emission into the radiation modes with outputs in the dielectric and the vacuum, respectively. The functions $F_{\mathrm{rad}}^{\mathrm{mat}}$ and $F_{\mathrm{rad}}^{\mathrm{vac}}$ are determined as the results of the application of the transformation $\mathcal{T}=(\mathbf{u}\to\mathbf{u}^*,\phi\to\phi+\pi)$ to the functions $F_{\mathrm{rad}}^{(1)}$ and $F_{\mathrm{rad}}^{(2)}$, respectively. 
Here, we have introduced the notations $F_{\mathrm{rad}}^{(1)}=F_{\mathrm{rad}}^{s1}+F_{\mathrm{rad}}^{p1}$ and $F_{\mathrm{rad}}^{(2)}=F_{\mathrm{rad}}^{s2}+F_{\mathrm{rad}}^{p2}$
for the angular densities of the spontaneous emission rates into the radiation modes with single inputs incident from medium 1 and medium 2 to the interface, respectively.
When we perform the above-described procedure, we get
\begin{eqnarray}\label{fl62}
\lefteqn{F_{\mathrm{rad}}^{\mathrm{mat}}=\frac{3}{8\pi\xi} (1- r_s^2) \big[|u_y|^2\sin^2\phi+|u_z|^2\cos^2\phi}\nonumber\\
&&\mbox{} -\mathrm{Re\,}(u_y^*u_z)\sin2\phi\big]\nonumber\\       
&&\mbox{} +\frac{3}{8\pi\xi}\big\{(1-r_p^2)[|u_x|^2(1-\xi^2)+|u_y|^2\xi^2\cos^2\phi \nonumber\\
&&\mbox{}  +|u_z|^2\xi^2\sin^2\phi + \mathrm{Re\,}(u_y^*u_z)\xi^2\sin2\phi] \nonumber\\
&&\mbox{}  + 2(1-r_p^2)\xi\sqrt{1-\xi^2}\,\mathrm{Re\,}(u_x^*u_y\cos\phi + u_x^*u_z\sin\phi)\big\}\qquad
\end{eqnarray}
and
\begin{eqnarray}\label{fl63}
\lefteqn{F_{\mathrm{rad}}^{\mathrm{vac}}=\frac{3}{8\pi\xi}\big[1+r_s^2+2r_s\cos(2\xi k_0x)\big] \big[|u_y|^2\sin^2\phi} \nonumber\\
&&\mbox{} +|u_z|^2\cos^2\phi -\mathrm{Re\,}(u_y^*u_z)\sin2\phi\big] \nonumber\\       
&&\mbox{} +\frac{3}{8\pi\xi}\big\{(1+r_p^2)[|u_x|^2(1-\xi^2)+|u_y|^2\xi^2\cos^2\phi \nonumber\\
&&\mbox{}  +|u_z|^2\xi^2\sin^2\phi + \mathrm{Re\,}(u_y^*u_z)\xi^2\sin2\phi] \nonumber\\
&&\mbox{}  - 2(1-r_p^2)\xi\sqrt{1-\xi^2}\,\mathrm{Re\,}(u_x^*u_y\cos\phi + u_x^*u_z\sin\phi)\big\}\nonumber\\
&&\mbox{}  + \frac{3}{4\pi\xi}r_p \big\{\cos(2\xi k_0x)\big[|u_x|^2(1-\xi^2)-|u_y|^2\xi^2\cos^2\phi \nonumber\\
&&\mbox{} - |u_z|^2\xi^2\sin^2\phi  - \mathrm{Re\,}(u_y^*u_z)\xi^2\sin2\phi\big] \nonumber\\
&&\mbox{} + 2\xi\sqrt{1-\xi^2}\sin(2\xi k_0x)\nonumber\\
&&\mbox{} \times \mathrm{Im\,}(u_x^*u_y\cos\phi + u_x^*u_z\sin\phi) \big\}.
\end{eqnarray}
According to Eq.~\eqref{fl62}, the angular density $F_{\mathrm{rad}}^{\mathrm{mat}}$ of the rate of spontaneous emission into the radiation modes with outputs in the dielectric does not depend on the atom-interface distance $x$.

The differences $\Delta F_{\mathrm{rad}}^{\mathrm{mat}}\equiv F_{\mathrm{rad}}^{\mathrm{mat}}(\xi,\phi)-F_{\mathrm{rad}}^{\mathrm{mat}}(\xi,\phi+\pi)$ and $\Delta F_{\mathrm{rad}}^{\mathrm{vac}}\equiv F_{\mathrm{rad}}^{\mathrm{vac}}(\xi,\phi)-F_{\mathrm{rad}}^{\mathrm{vac}}(\xi,\phi+\pi)$ are found from Eqs.~\eqref{fl62} and \eqref{fl63} to be 
\begin{equation}\label{fl67}
\begin{split}
\Delta F_{\mathrm{rad}}^{\mathrm{mat}}&=
\frac{3}{2\pi}\sqrt{1-\xi^2}\,(1-r_p^2)\\
&\quad\times\mathrm{Re\,}(u_x^*u_y\cos\phi + u_x^*u_z\sin\phi)
\end{split}
\end{equation}
and
\begin{equation}\label{fl68}
\begin{split}
\Delta F_{\mathrm{rad}}^{\mathrm{vac}}&=
-\frac{3}{2\pi}\sqrt{1-\xi^2}\,(1-r_p^2)\\
&\quad\times\mathrm{Re\,}(u_x^*u_y\cos\phi + u_x^*u_z\sin\phi)\\
&\quad+\frac{3}{\pi}\sqrt{1-\xi^2}\,r_p\sin(2\xi k_0x)\\
&\quad\times\mathrm{Im\,}(u_x^*u_y\cos\phi + u_x^*u_z\sin\phi).
\end{split}
\end{equation}
It is clear that $\Delta F_{\mathrm{rad}}^{\mathrm{mat}}+\Delta F_{\mathrm{rad}}^{\mathrm{vac}}=\Delta F_{\mathrm{rad}}$, where $\Delta F_{\mathrm{rad}}$ is given by Eq.~\eqref{fl45}.

According to Eq.~\eqref{fl67}, the difference $\Delta F_{\mathrm{rad}}^{\mathrm{mat}}$ between the rate densities of spontaneous emission into the radiation modes outgoing into the dielectric
with opposite in-plane wave vectors does not depend on the atom-interface distance $x$. This difference is associated with the coefficients $\mathrm{Re\,}(u_x^*u_y)$ and $\mathrm{Re\,}(u_x^*u_z)$. It can be
nonzero when the atomic dipole polarization vector is a real vector tilted with respect to the axis $x$ and to the interface plane $yz$. Thus, $\Delta F_{\mathrm{rad}}^{\mathrm{mat}}$ is just the result of the geometric asymmetry of the orientation of the dipole vector with respect to the interface plane.

Equation \eqref{fl68} shows that
the difference $\Delta F_{\mathrm{rad}}^{\mathrm{vac}}$ for the radiation modes with outputs in the vacuum has two contributions, one is associated with the coefficient $1-r_p^2$ and the other one is associated with the coefficient $r_p$. The first contribution is equal to  $-\Delta F_{\mathrm{rad}}^{\mathrm{mat}}$ and is
caused by the asymmetry of the orientation of the dipole vector with respect to the interface plane.
The second contribution is equal to $\Delta F_{\mathrm{rad}}$ and is related to spin-orbit coupling of light \cite{Zeldovich,Bliokh review}.

We introduce the notations $\gamma_{\mathrm{rad}}^{\mathrm{mat}(+)}=\gamma_0\int_0^1 \xi d\xi\int_0^{\pi} F_{\mathrm{rad}}^{\mathrm{mat}} d\phi$ and $\gamma_{\mathrm{rad}}^{\mathrm{mat}(-)}=\gamma_0\int_0^1 \xi d\xi\int_{\pi}^{2\pi} F_{\mathrm{rad}}^{\mathrm{mat}} d\phi$ for the rates of spontaneous emission into the radiation modes outgoing into 
the $+z$ and $-z$ sides, respectively, of the dielectric  half-space and, similarly, 
the notations $\gamma_{\mathrm{rad}}^{\mathrm{vac}(+)}=\gamma_0\int_0^1 \xi d\xi\int_0^{\pi} F_{\mathrm{rad}}^{\mathrm{vac}} d\phi$ and $\gamma_{\mathrm{rad}}^{\mathrm{vac}(-)}=\gamma_0\int_0^1 \xi d\xi\int_{\pi}^{2\pi} F_{\mathrm{rad}}^{\mathrm{vac}} d\phi$ for the rates of spontaneous emission into the radiation modes outgoing into 
the $+z$ and $-z$ sides, respectively, of the vacuum half-space. We find
\begin{equation}\label{fl69}
\begin{split}
\gamma_{\mathrm{rad}}^{\mathrm{mat}(\pm)}&=\frac{\gamma_{\mathrm{rad}}^{\mathrm{mat}}}{2}\pm\frac{\Delta_{\mathrm{rad}}^{\mathrm{mat}}}{2},\\
\gamma_{\mathrm{rad}}^{\mathrm{vac}(\pm)}&=\frac{\gamma_{\mathrm{rad}}^{\mathrm{vac}}}{2}\pm\frac{\Delta_{\mathrm{rad}}^{\mathrm{vac}}}{2}.
\end{split}
\end{equation}
Here, we have introduced the notations \cite{Nano-Optics}
\begin{eqnarray}\label{fl64}
\gamma_{\mathrm{rad}}^{\mathrm{mat}}&=&\frac{\gamma_0}{2}-\frac{3}{8}\gamma_0\int\limits_{0}^{1}\{(1-|u_x|^2)r_s^2(\xi) \nonumber\\
&&\mbox{} +[|u_x|^2(2-3\xi^2)+\xi^2]r_p^2(\xi)\}\,d\xi
\end{eqnarray}
and
\begin{eqnarray}\label{fl65}
\gamma_{\mathrm{rad}}^{\mathrm{vac}}&=&\frac{\gamma_0}{2} +\frac{3}{8}\gamma_0\int\limits_{0}^{1}\{(1-|u_x|^2)r_s^2(\xi)\nonumber\\ 
&&\mbox{} +[|u_x|^2(2-3\xi^2)+\xi^2]r_p^2(\xi)\}\,d\xi \nonumber\\
&&\mbox{} + \frac{3}{4}\gamma_0\int\limits_{0}^{1}\{(1-|u_x|^2)r_s(\xi)\nonumber\\
&&\mbox{} + [|u_x|^2(2-\xi^2)-\xi^2]r_p(\xi)\}\cos(2\xi k_0x)\,d\xi\qquad
\end{eqnarray}
for the rates of spontaneous emission into the radiation modes with outputs in the dielectric and the vacuum, respectively. We have also introduced the notations
\begin{eqnarray}\label{fl71}
\Delta_{\mathrm{rad}}^{\mathrm{mat}}&=&
\frac{1}{\pi}\gamma_0\,\mathrm{Re\,}(u_x^*u_z)\bigg[1- 3\int\limits_0^1\xi\sqrt{1-\xi^2}\,r_p^2(\xi)d\xi \bigg]\qquad
\end{eqnarray}
and
\begin{eqnarray}\label{fl72}
\lefteqn{\Delta_{\mathrm{rad}}^{\mathrm{vac}}=
-\frac{1}{\pi}\gamma_0\,\mathrm{Re\,}(u_x^*u_z)\bigg[1-3\int\limits_0^1\xi\sqrt{1-\xi^2}r_p^2(\xi)d\xi\bigg]}\nonumber\\
&&\mbox{}+\frac{6}{\pi}\gamma_0\,\mathrm{Im\,}(u_x^*u_z)\int\limits_{0}^{1}\xi\sqrt{1-\xi^2}\,r_p(\xi)\sin(2\xi k_0x)\,d\xi\qquad
\end{eqnarray}
for the differences between the rate components for the opposite sides $\pm z$ of the dielectric and the vacuum, respectively.
It is clear that $\gamma_{\mathrm{rad}}^{\mathrm{mat}(\pm)}$, $\gamma_{\mathrm{rad}}^{\mathrm{mat}}$, and $\Delta_{\mathrm{rad}}^{\mathrm{mat}}$ do not depend on  the atom-interface distance $x$ \cite{Nano-Optics}, while $\gamma_{\mathrm{rad}}^{\mathrm{vac}(\pm)}$, $\gamma_{\mathrm{rad}}^{\mathrm{vac}}$, and $\Delta_{\mathrm{rad}}^{\mathrm{vac}}$ oscillate with increasing $x$.

We now derive the radiation patterns of spontaneous emission into radiation modes with outputs on a given side of the interface in the far-field limit. For the radiation modes with outputs in the half-space $x<0$, the angle $\theta$ between the wave vector $(\beta_1,K_y,K_z)$ and the axis $x$ is given by the formulas $n_1\sin\theta=\kappa=\sqrt{1-\xi^2}$ and $n_1\cos\theta=-\eta$
for $\theta\in[\pi-\arcsin(1/n_1),\pi]$. 
For the radiation modes with outputs in the half-space $x>0$, the angle $\theta$ between the wave vector $(\beta_2,K_y,K_z)$ and the axis $x$ is given by the formulas $\sin\theta=\kappa=\sqrt{1-\xi^2}$ and $\cos\theta=\xi$ for $\theta\in[0,\pi/2]$. 
Then, we find
$F_{\mathrm{rad}}^{\mathrm{mat}}(\xi,\phi)\xi d\xi d\phi=-P_{\mathrm{rad}}^{\mathrm{mat}}(\theta,\phi)\sin\theta d\theta d\phi$ 
and
$F_{\mathrm{rad}}^{\mathrm{vac}}(\xi,\phi)\xi d\xi d\phi=P_{\mathrm{rad}}^{\mathrm{vac}}(\theta,\phi)\sin\theta d\theta d\phi$, 
where 
\begin{equation}\label{fl66}
\begin{split}
P_{\mathrm{rad}}^{\mathrm{mat}}&=n_1\eta F_{\mathrm{rad}}^{\mathrm{mat}}=-n_1^2\cos\theta F_{\mathrm{rad}}^{\mathrm{mat}},\\
P_{\mathrm{rad}}^{\mathrm{vac}}&=\xi F_{\mathrm{rad}}^{\mathrm{vac}}=\cos\theta F_{\mathrm{rad}}^{\mathrm{vac}}.
\end{split}
\end{equation}
The functions $P_{\mathrm{rad}}^{\mathrm{mat}}$ and $P_{\mathrm{rad}}^{\mathrm{vac}}$
are the angular distributions of spontaneous emission into radiation modes with respect to the spherical angles $\theta$ and $\phi$. 
In the particular case where the dipole polarization vector $\mathbf{u}$ is real, the expressions for $P_{\mathrm{rad}}^{\mathrm{mat}}$ and $P_{\mathrm{rad}}^{\mathrm{vac}}$ reduce to the results for the far-field limit of the radiation patterns in the allowed region inside and the half-space outside the dielectric medium, respectively \cite{Nano-Optics}.

\section{Numerical results}
\label{sec:numerical}

We perform numerical calculations. For the wavelength of the atomic transition, we use the value $\lambda_0=852$~nm corresponding to the $D_2$ line of atomic cesium.
For the refractive index of the dielectric medium, we use the value $n_1=1.45$ corresponding to silica.

\begin{figure}[tbh]
\begin{center}
\includegraphics{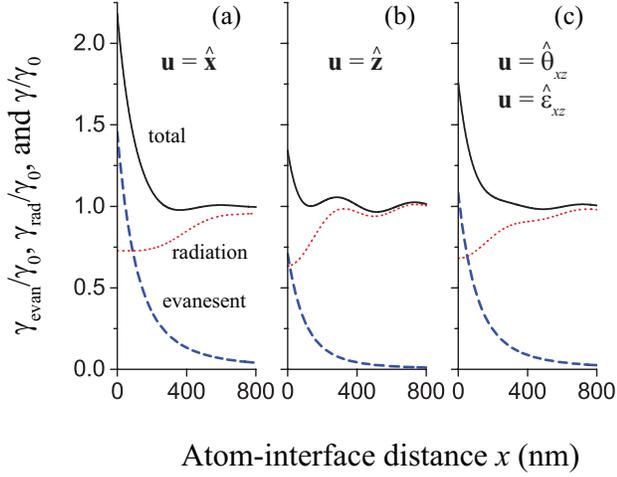}
\end{center}
\caption{(Color online) 
Rates $\gamma_{\mathrm{evan}}$, $\gamma_{\mathrm{rad}}$, and $\gamma$ of spontaneous emission from a two-level atom  
into evanescent modes (dashed blue lines), radiation modes (dotted red lines), and  both types of modes (solid black lines), respectively, as functions of the atom-interface distance $x$. The atomic dipole polarization vector $\mathbf{u}$  is equal to $\hat{\mathbf{x}}$ (a), $\hat{\mathbf{z}}$ (b), and $\hat{\boldsymbol{\theta}}_{xz}\equiv(\hat{\mathbf{x}}+\hat{\mathbf{z}})/\sqrt{2}$ 
or $\hat{\boldsymbol{\varepsilon}}_{xz}\equiv(\hat{\mathbf{x}}+i\hat{\mathbf{z}})/\sqrt{2}$ (c).
The rates are normalized to the spontaneous emission rate $\gamma_0$ of the atom in free space. The refractive index of the medium is $n_1=1.45$. The wavelength of the atomic transition is $\lambda_0=852$ nm.
}
\label{fig2}
\end{figure}

According to the previous section, the rates $\gamma_{\mathrm{evan}}$, $\gamma_{\mathrm{rad}}$, and $\gamma$ of spontaneous emission from a two-level atom into evanescent modes, radiation modes, and  both types of modes, respectively, are determined by Eqs.~\eqref{fl33}, \eqref{fl54}, and \eqref{fl60}, respectively. We plot in Fig.~\ref{fig2} the normalized rates $\gamma_{\mathrm{evan}}/\gamma_0$, $\gamma_{\mathrm{rad}}/\gamma_0$, and $\gamma/\gamma_0$ as functions of the atom-interface distance $x$. Figures \ref{fig2}(a) and \ref{fig2}(b) correspond respectively to the cases where the dipole polarization vector $\mathbf{u}$ is equal to $\hat{\mathbf{x}}$ and $\hat{\mathbf{z}}$. The results for the cases where $\mathbf{u}=\hat{\boldsymbol{\theta}}_{xz}\equiv(\hat{\mathbf{x}}+\hat{\mathbf{z}})/\sqrt{2}$ and $\mathbf{u}=\hat{\boldsymbol{\varepsilon}}_{xz}\equiv(\hat{\mathbf{x}}+i\hat{\mathbf{z}})/\sqrt{2}$ are the same and are shown in Fig.~\ref{fig2}(c). The solid black curves for the normalized total rate $\gamma/\gamma_0$ show not only the enhancement, $\gamma/\gamma_0>1$, but also the inhibition, $\gamma/\gamma_0<1$, of spontaneous emission, depending on the atom-interface distance $x$. Such changes are quantum electrodynamic effects resulting from modifications of the field mode structure in the presence of the dielectric  \cite{Agarwal,Wylie,Lukosz}. The enhancement of the total rate of spontaneous emission, $\gamma/\gamma_0>1$, is mainly due to the presence of emission into evanescent modes. The maximum value of $\gamma/\gamma_0$ is about $2.18$, achieved at $x=0$ for $\mathbf{u}=\hat{\mathbf{x}}$. We observe a rapid decrease of $\gamma_{\mathrm{evan}}$ and oscillations of $\gamma_{\mathrm{rad}}$ and $\gamma$ as $x$ increases. The rapid decrease of $\gamma_{\mathrm{evan}}$ is a consequence of the tight confinement of evanescent modes in the direction $+x$. The oscillations of $\gamma_{\mathrm{rad}}$ and $\gamma$ are due to the interference between the emitted and reflected fields. The period of oscillations is roughly equal to one half of the wavelength $\lambda_0$ of the atomic transition [see Eqs.~\eqref{fl54} and \eqref{fl60}].  The dotted red curves in Fig.~\ref{fig2} show that the interference is destructive, $\gamma_{\mathrm{rad}}/\gamma_0<1$, when the atom is close to the interface, and may become constructive, $\gamma_{\mathrm{rad}}/\gamma_0>1$, in some specific regions where the atom is not too close to the interface. The inhibition of the total spontaneous emission, $\gamma/\gamma_0<1$, may occur in some specific regions of $x$. In the limit of large distance $x$, we have $\gamma_{\mathrm{evan}}\to0$ and $\gamma\to\gamma_{\mathrm{rad}}\to\gamma_0$.

\begin{figure}[tbh]
\begin{center}
\includegraphics{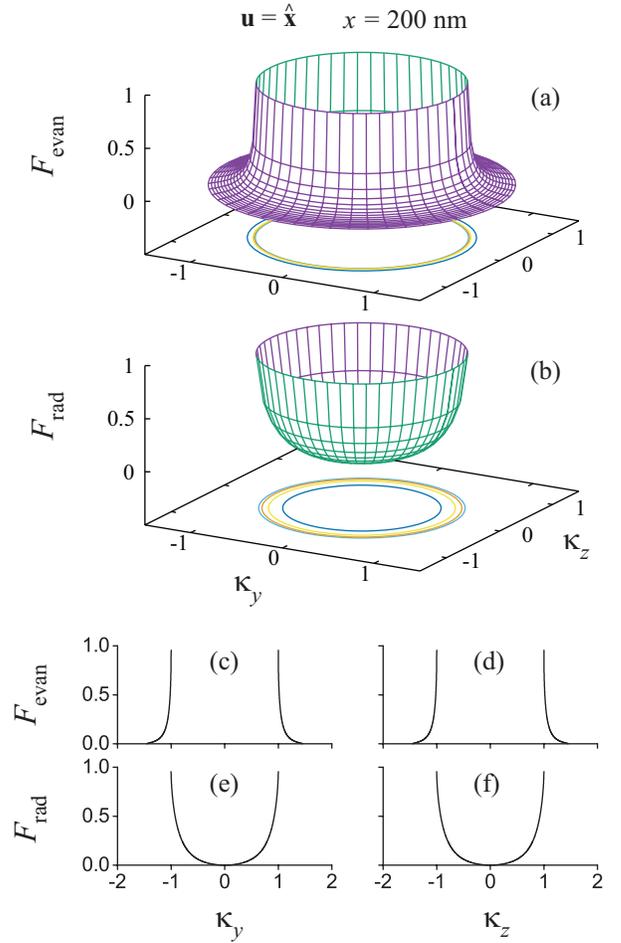}
\end{center}
\caption{(Color online) Angular densities $F_{\mathrm{evan}}$ (a) and  $F_{\mathrm{rad}}$ (b) of the rates of spontaneous emission into evanescent and radiation modes, respectively, as functions of $\kappa_y$ and $\kappa_z$ in the case where the dipole polarization vector $\mathbf{u}$ is aligned along the axis $x$ and the atom-interface distance is $x=200$ nm. Other parameters are as for Fig.~\ref{fig2}. 
The contour lines of the surface plots are shown to help visualization. The bottom panel shows the one-dimensional profiles of $F_{\mathrm{evan}}$  and  $F_{\mathrm{rad}}$. In (c) and (e), $\kappa_z=0$. In (d) and (f), $\kappa_y=0$. 
}
\label{fig3}
\end{figure}

\begin{figure}[tbh]
\begin{center}
\includegraphics{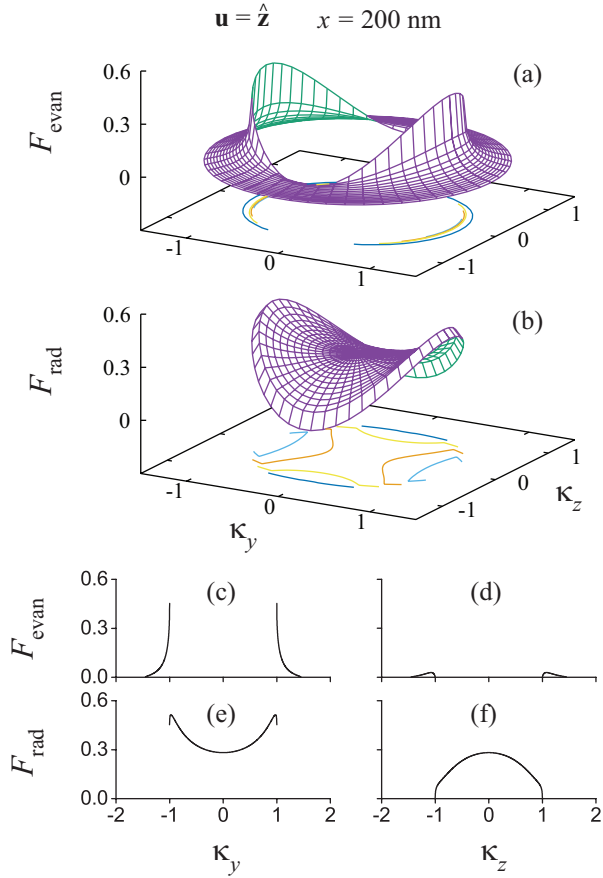} 
\end{center}
\caption{(Color online) 
Angular densities $F_{\mathrm{evan}}$ and $F_{\mathrm{rad}}$ of the rates of spontaneous emission into evanescent and radiation modes, respectively, in the case where the dipole polarization vector $\mathbf{u}$ is aligned along the axis $z$ and the atom-interface distance is $x=200$ nm. Other parameters are as for Fig.~\ref{fig3}.
}
\label{fig4}
\end{figure}

\begin{figure}[tbh]
\begin{center}
\includegraphics{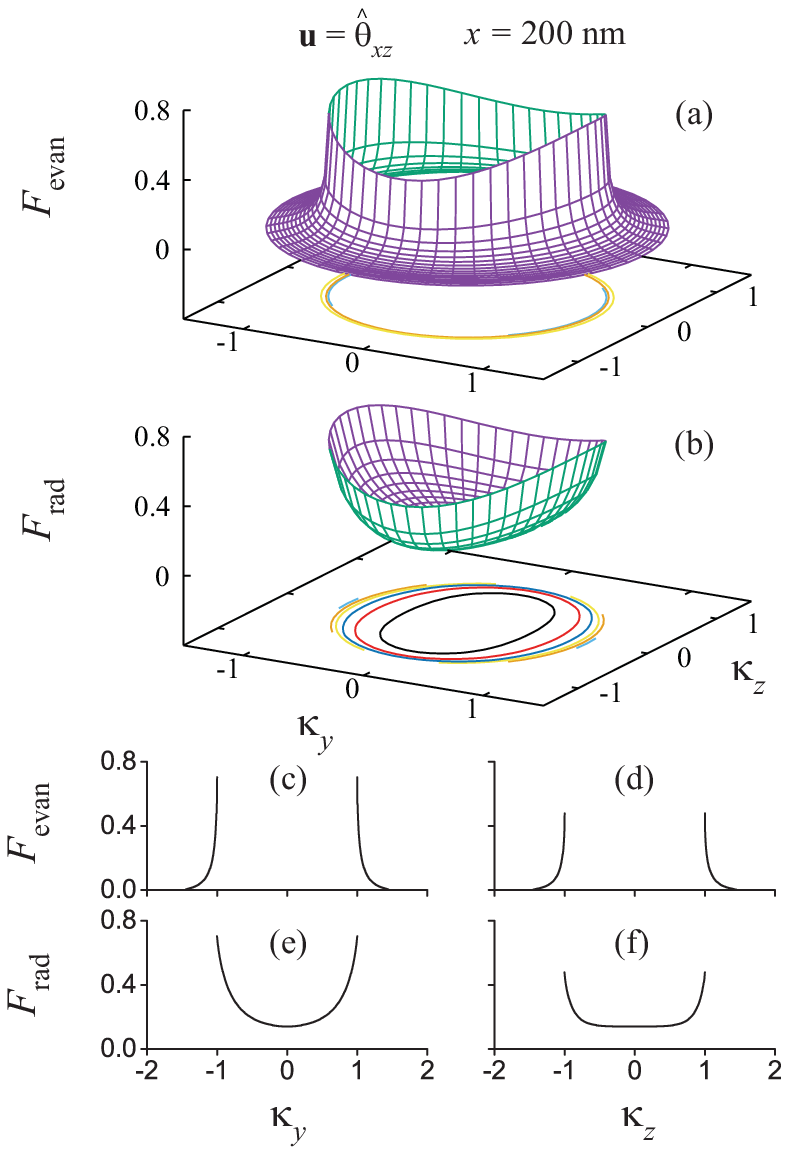}
\end{center}
\caption{(Color online) 
Angular densities $F_{\mathrm{evan}}$ and  $F_{\mathrm{rad}}$ of the rates of spontaneous emission into evanescent and radiation modes, respectively, in the case where the dipole polarization vector is $\mathbf{u}=\hat{\boldsymbol{\theta}}_{xz}\equiv (\hat{\mathbf{x}}+\hat{\mathbf{z}})/\sqrt{2}$ and the atom-interface distance is $x=200$ nm. Other parameters are as for Fig.~\ref{fig3}. 
}
\label{fig5}
\end{figure}

\begin{figure}[tbh]
\begin{center}
\includegraphics{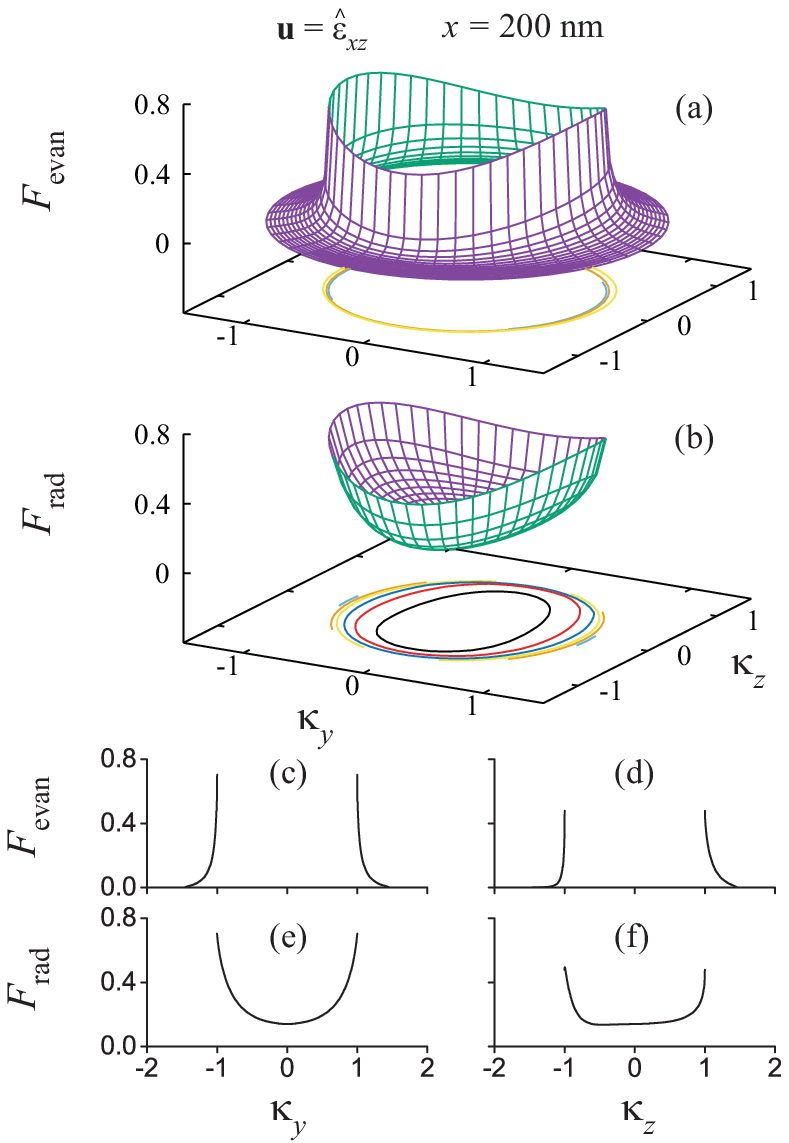}
\end{center}
\caption{(Color online) 
Angular densities $F_{\mathrm{evan}}$ and  $F_{\mathrm{rad}}$ of the rates of spontaneous emission into evanescent and radiation modes, respectively, in the case where the dipole polarization vector is $\mathbf{u}=\hat{\boldsymbol{\varepsilon}}_{xz}\equiv (\hat{\mathbf{x}}+i\hat{\mathbf{z}})/\sqrt{2}$ and the atom-interface distance is $x=200$ nm. Other parameters are as for Fig.~\ref{fig3}. 
}
\label{fig6}
\end{figure}

According to the previous section, the angular densities $F_{\mathrm{evan}}$ and $F_{\mathrm{rad}}$ of the rates of spontaneous emission into evanescent and radiation modes, respectively, are given by Eqs.~\eqref{fl17a} and \eqref{fl40a}, respectively. We plot in Figs.~\ref{fig3}--\ref{fig6} the angular densities $F_{\mathrm{evan}}$ and $F_{\mathrm{rad}}$ as functions of the components $\kappa_y$ and $\kappa_z$ of the normalized in-plane wave vector $\boldsymbol{\kappa}=(0,\kappa_y,\kappa_z)=\mathbf{K}/k_0=(0,K_y,K_z)/k_0$. 
The dipole polarization vector $\mathbf{u}$ is chosen to be equal to $\hat{\mathbf{x}}$ (Fig.~\ref{fig3}), $\hat{\mathbf{z}}$ (Fig.~\ref{fig4}), $\hat{\boldsymbol{\theta}}_{xz}$ (Fig.~\ref{fig5}), and $\hat{\boldsymbol{\varepsilon}}_{xz}$ (Fig.~\ref{fig6}). The distance from the atom to the interface is $x=200$ nm. 

We observe that in the case of Fig.~\ref{fig3}, where $\mathbf{u}$ is aligned along the axis $x$, the angular densities $F_{\mathrm{evan}}$ and $F_{\mathrm{rad}}$ are cylindrically symmetric functions of $\boldsymbol{\kappa}$. In the cases of Fig.~\ref{fig4}, where $\mathbf{u}$ is aligned along the axis $z$, and Fig.~\ref{fig5}, where $\mathbf{u}$ is aligned at a nonzero angle with respect to the axis $x$ in the $xz$ plane, $F_{\mathrm{evan}}$ and  $F_{\mathrm{rad}}$ are not cylindrically symmetric but are symmetric under the transformations $\kappa_y\to -\kappa_y$ and $\kappa_z\to -\kappa_z$. Thus, in the cases of Figs.~\ref{fig3}--\ref{fig5},
where $\mathbf{u}$ is a real vector, $F_{\mathrm{evan}}$ and  $F_{\mathrm{rad}}$ are symmetric under the transformation 
$\boldsymbol{\kappa}\to -\boldsymbol{\kappa}$.  

In the case of Fig.~\ref{fig6}, where $\mathbf{u}$ is a complex vector, that is, where the atomic dipole rotates with time in the $xz$ plane, $F_{\mathrm{evan}}$ and  $F_{\mathrm{rad}}$ are symmetric under the transformation $\kappa_y\to -\kappa_y$ [see Figs.~\ref{fig6}(c) and \ref{fig6}(e)] but not symmetric under the transformation $\kappa_z\to -\kappa_z$ [see Figs.~\ref{fig6}(d) and \ref{fig6}(f)] and, consequently, not symmetric under the transformation $\boldsymbol{\kappa}\to -\boldsymbol{\kappa}$. The asymmetry between the rates for the opposite in-plane wave vectors $\mathbf{K}$ and $-\mathbf{K}$ results from the overlap between the ellipticity vector of the dipole polarization of the atom and the ellipticity vector of the local electric polarization of the field mode. Figures \ref{fig3}--\ref{fig6} show that, in the limit $\kappa\to1$, the angular densities $F_{\mathrm{evan}}$ and $F_{\mathrm{rad}}$ approach the same limiting values and there is no difference between the limiting values of the rates for the modes with the opposite in-plane wave vectors $\mathbf{K}$ and $-\mathbf{K}$.
These numerical results are in agreement with the analytical results of the previous section.

\begin{figure}[tbh]
\begin{center}
\includegraphics{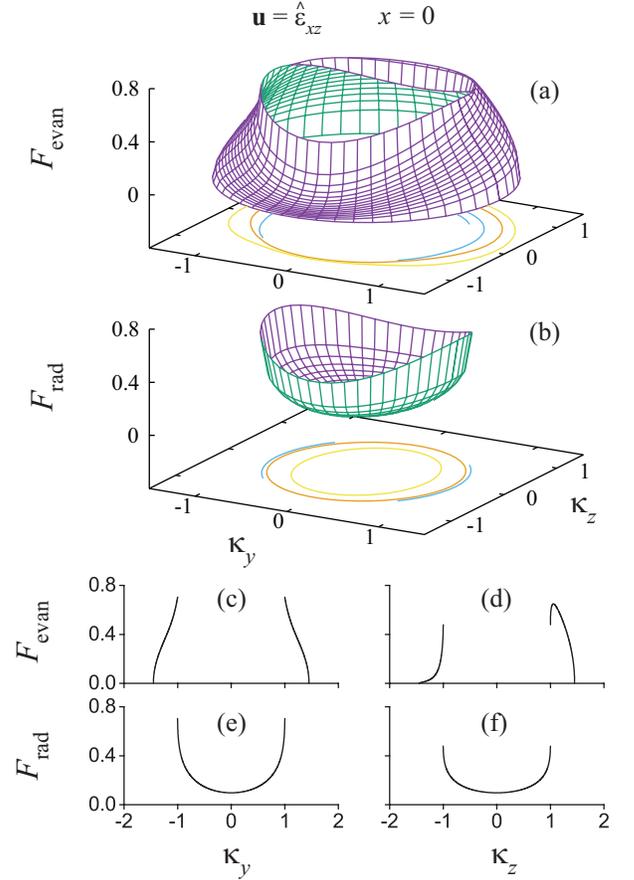}
\end{center}
\caption{(Color online) 
Same as Fig.~\ref{fig6} but the distance from the atom to the interface is $x=0$.
}
\label{fig7}
\end{figure}

\begin{figure}[tbh]
\begin{center}
\includegraphics{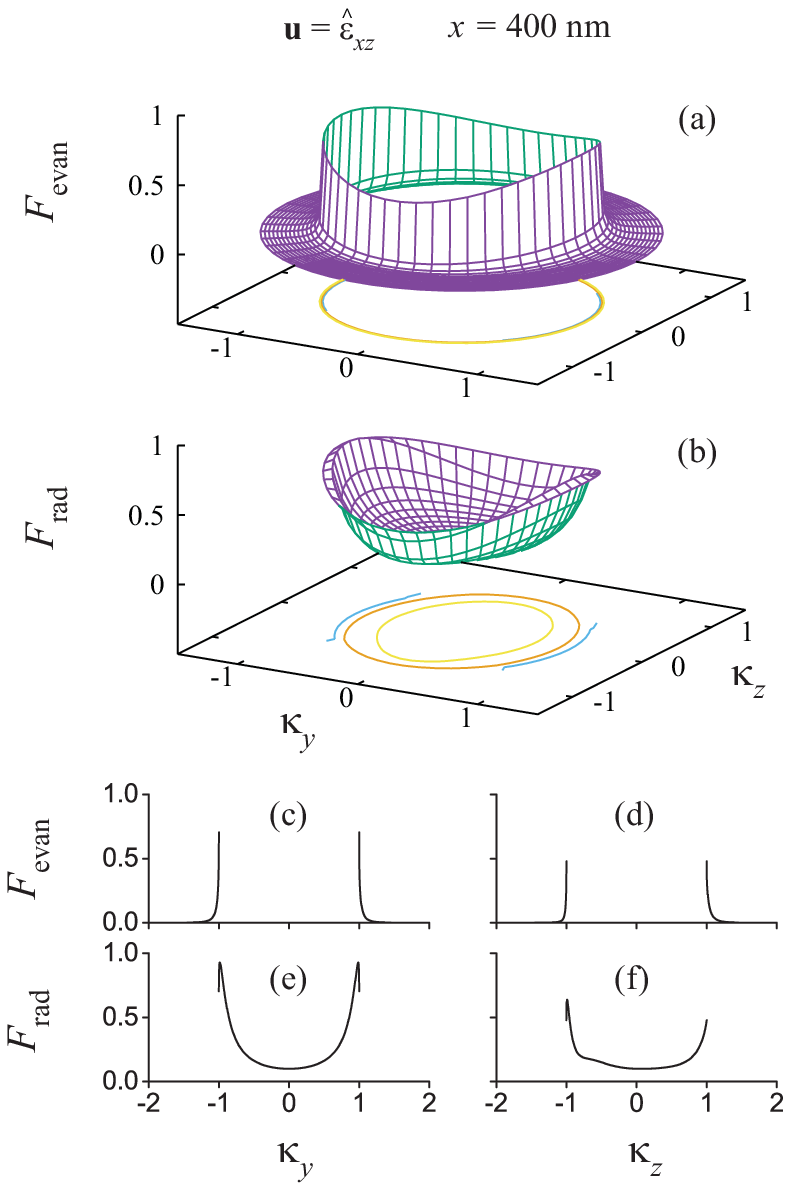}
\end{center}
\caption{(Color online) 
Same as Fig.~\ref{fig6} but the distance from the atom to the interface is $x=400$ nm.
}
\label{fig8}
\end{figure}

\begin{figure}[tbh]
\begin{center}
\includegraphics{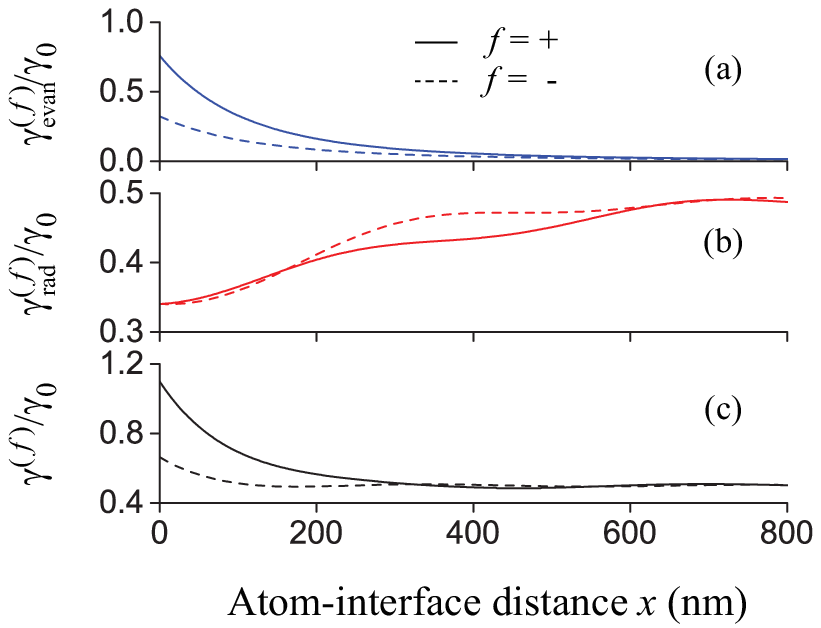}
\end{center}
\caption{(Color online) Rates $\gamma_{\mathrm{evan}}^{(f)}$ (a), $\gamma_{\mathrm{rad}}^{(f)}$ (b), and $\gamma^{(f)}$ (c) of spontaneous emission into evanescent modes, radiation modes, and both types of modes, respectively, propagating into the positive side $f=+$ (solid lines) or negative side $f=-$ (dashed lines) of the axis $z$
as functions of the atom-interface distance $x$. The polarization vector of the atomic dipole is $\mathbf{u}=\hat{\boldsymbol{\varepsilon}}_{xz}$.
The rates are normalized to the spontaneous emission rate $\gamma_0$ of the atom in free space. Other parameters are as for Fig.~\ref{fig2}.
}
\label{fig9}
\end{figure}

\begin{figure}[tbh]
\begin{center}
\includegraphics{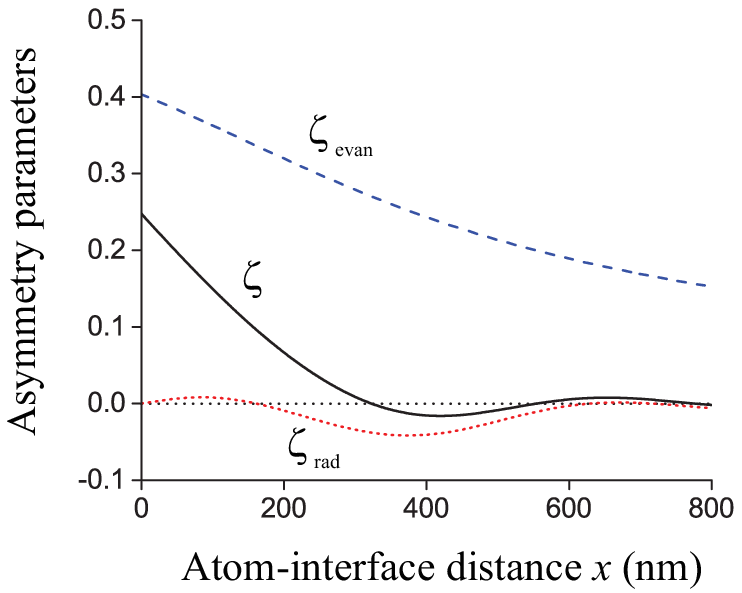}
\end{center}
\caption{(Color online) Asymmetry parameters $\zeta_{\mathrm{evan}}$ (dashed blue line), $\zeta_{\mathrm{rad}}$ (dotted red line), and $\zeta$ (solid black line) for the rates of directional spontaneous emission into evanescent modes, radiation modes, and both types of modes, respectively, as functions of the atom-interface distance $x$. The polarization vector of the atomic dipole is $\mathbf{u}=\hat{\boldsymbol{\varepsilon}}_{xz}$. Other parameters are as for Fig.~\ref{fig2}. The short-dotted black line is for the zero value of the asymmetry parameters and is a guide to the eye.
}
\label{fig10}
\end{figure}

In Figs.~\ref{fig7}--\ref{fig10}, we study in more detail the case $\mathbf{u}=\hat{\boldsymbol{\varepsilon}}_{xz}$. We focus on this case in order to get insight into
the asymmetry of the angular distributions $F_{\mathrm{evan}}$ and  $F_{\mathrm{rad}}$ with respect to central inversion in the interface plane.

In order to see the effect of the atom-interface distance $x$ on the asymmetry of spontaneous emission, we plot 
in Figs.~\ref{fig7} and \ref{fig8} the angular densities of the rates of spontaneous emission from an atom with the dipole polarization vector $\mathbf{u}=\hat{\boldsymbol{\varepsilon}}_{xz}$ at the distances
$x=0$ and $x=400$ nm, respectively. Other parameters are as for Fig.~\ref{fig6}. 

We observe from Fig.~\ref{fig7} that, when $x=0$, the angular density $F_{\mathrm{evan}}$ of the rate of spontaneous emission into evanescent modes is strongly asymmetric with respect to the transformation 
$\kappa_z\to -\kappa_z$ and, hence, the transformation $\boldsymbol{\kappa}\to -\boldsymbol{\kappa}$ [see Figs.~\ref{fig7}(a) and \ref{fig7}(d)], while the angular density $F_{\mathrm{rad}}$ of the rate of spontaneous emission into radiation modes is symmetric [see Figs.~\ref{fig7}(b) and \ref{fig7}(f)]. Comparison between Figs.~\ref{fig8}(a) and \ref{fig7}(a) shows that the density of the rate of spontaneous emission into evanescent modes in the case of Fig.~\ref{fig8}(a), where $x=400$ nm, reduces with increasing $\kappa$ much faster than that in the case of Fig.~\ref{fig7}(a), where $x=0$.

According to the previous section, the rates $\gamma_{\mathrm{evan}}^{(f)}$, $\gamma_{\mathrm{rad}}^{(f)}$, and $\gamma^{(f)}$ of spontaneous emission into evanescent modes, radiation modes, and both types of modes, respectively, propagating into the side $f=+,-$ of the axis $z$, are determined by Eqs.~\eqref{fl32}, \eqref{fl53}, and \eqref{fl59}, respectively. We plot in Fig.~\ref{fig9} the rates $\gamma_{\mathrm{evan}}^{(f)}$, $\gamma_{\mathrm{rad}}^{(f)}$, and $\gamma^{(f)}$ as functions of the atom-interface distance $x$ in the case of $\mathbf{u}=\hat{\boldsymbol{\varepsilon}}_{xz}$. Figure \ref{fig9}(a) shows that the rates $\gamma_{\mathrm{evan}}^{(+)}$ and $\gamma_{\mathrm{evan}}^{(-)}$ of directional spontaneous emission into evanescent modes quickly decrease to zero with increasing $x$ and the inequality $\gamma_{\mathrm{evan}}^{(+)}>\gamma_{\mathrm{evan}}^{(-)}$
holds true for every $x\ge0$. Meanwhile, Fig. \ref{fig9}(b) shows that the rates $\gamma_{\mathrm{rad}}^{(+)}$ and $\gamma_{\mathrm{rad}}^{(-)}$ of directional spontaneous emission into radiation modes oscillate with increasing $x$ and approach the value $\gamma_0/2$ in the limit $x\to+\infty$. We observe that the equality $\gamma_{\mathrm{rad}}^{(+)}=\gamma_{\mathrm{rad}}^{(-)}$ holds true for $x=0$ and that both inequalities $\gamma_{\mathrm{rad}}^{(+)}>\gamma_{\mathrm{rad}}^{(-)}$ and $\gamma_{\mathrm{rad}}^{(+)}<\gamma_{\mathrm{rad}}^{(-)}$ are possible depending on the distance $x$.

The asymmetries between the rates $\gamma_{\mathrm{evan}}^{(+)}$ and $\gamma_{\mathrm{evan}}^{(-)}$, between the rates $\gamma_{\mathrm{rad}}^{(+)}$ and $\gamma_{\mathrm{rad}}^{(-)}$, and between the rates $\gamma^{(+)}$ and $\gamma^{(-)}$ are, as already stated in the previous section, characterized by the parameters $\zeta_{\mathrm{evan}}=\Delta_{\mathrm{evan}}/\gamma_{\mathrm{evan}}$, $\zeta_{\mathrm{rad}}=\Delta_{\mathrm{rad}}/\gamma_{\mathrm{rad}}$, and $\zeta=\Delta/\gamma$, respectively. We plot in Fig.~\ref{fig10} the asymmetry parameters $\zeta_{\mathrm{evan}}$, $\zeta_{\mathrm{rad}}$, and $\zeta$ as functions of the atom-interface distance $x$ in the case of $\mathbf{u}=\hat{\boldsymbol{\varepsilon}}_{xz}$. The dashed blue curve of the figure shows that the asymmetry parameter $\zeta_{\mathrm{evan}}$ for emission into evanescent modes is positive and monotonically decreases with increasing $x$. The dotted red and solid black curves of the figure show that the asymmetry parameters $\zeta_{\mathrm{rad}}$ and $\zeta$ for emission into radiation modes and both types of modes, respectively, oscillate with increasing $x$ and can be positive or negative depending on the distance $x$. For $x=0$, we have $\zeta_{\mathrm{rad}}=0$ and $\zeta_{\mathrm{evan}}>\zeta>0$. In the limit of large $x$, we have $\zeta\simeq\zeta_{\mathrm{rad}}\simeq0$.
In this limit, $\zeta_{\mathrm{evan}}$ is also small.

\begin{figure}[tbh]
\begin{center}
\includegraphics{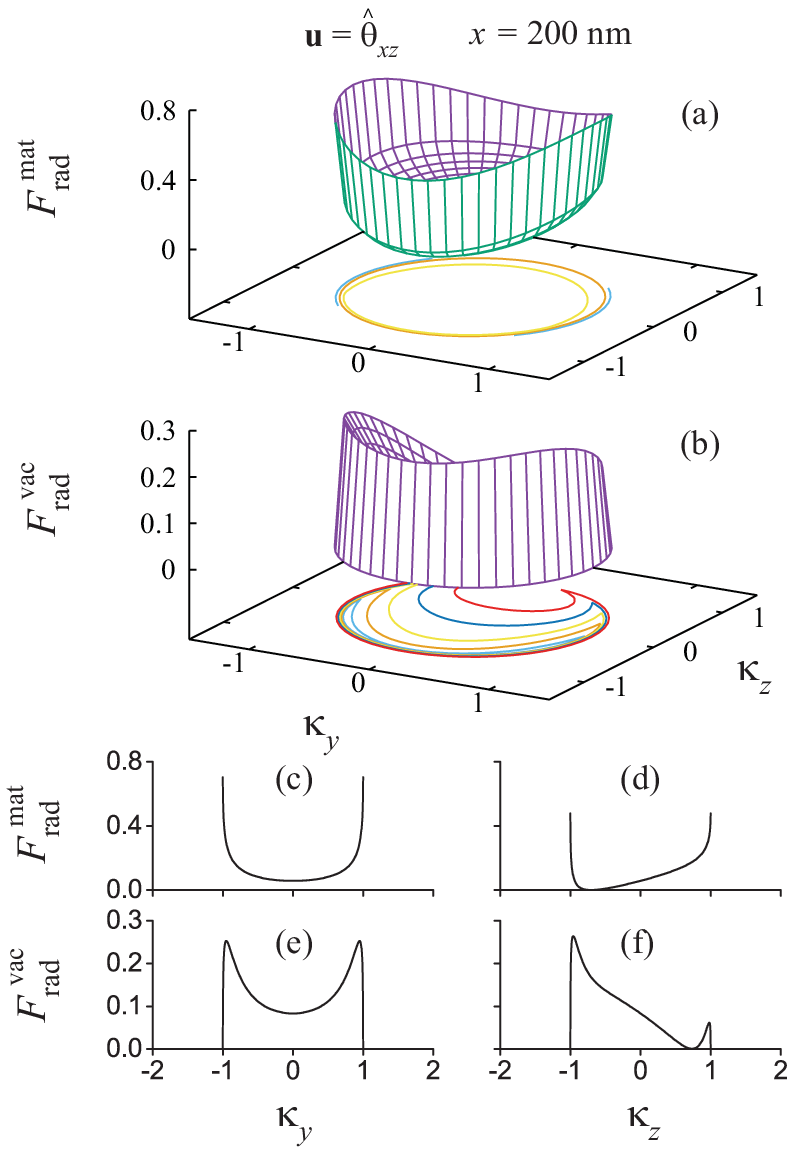}
\end{center}
\caption{(Color online) Angular densities $F_{\mathrm{rad}}^{\mathrm{mat}}$ (a) and  $F_{\mathrm{rad}}^{\mathrm{vac}}$ (b) of the rates of spontaneous emission into radiation modes
with the outputs inside and outside  the dielectric, respectively, as functions of $\kappa_y$ and $\kappa_z$ in the case of Fig.~\ref{fig5}, where 
the dipole polarization vector is $\mathbf{u}=\hat{\boldsymbol{\theta}}_{xz}$ and the atom-interface distance is $x=200$ nm.
Other parameters are as for Fig.~\ref{fig2}. 
The bottom panel shows the one-dimensional profiles of $F_{\mathrm{rad}}^{\mathrm{mat}}$ and $F_{\mathrm{rad}}^{\mathrm{vac}}$. 
In (c) and (e), $\kappa_z=0$. In (d) and (f), $\kappa_y=0$. 
}
\label{fig11}
\end{figure}

\begin{figure}[tbh]
\begin{center}
\includegraphics{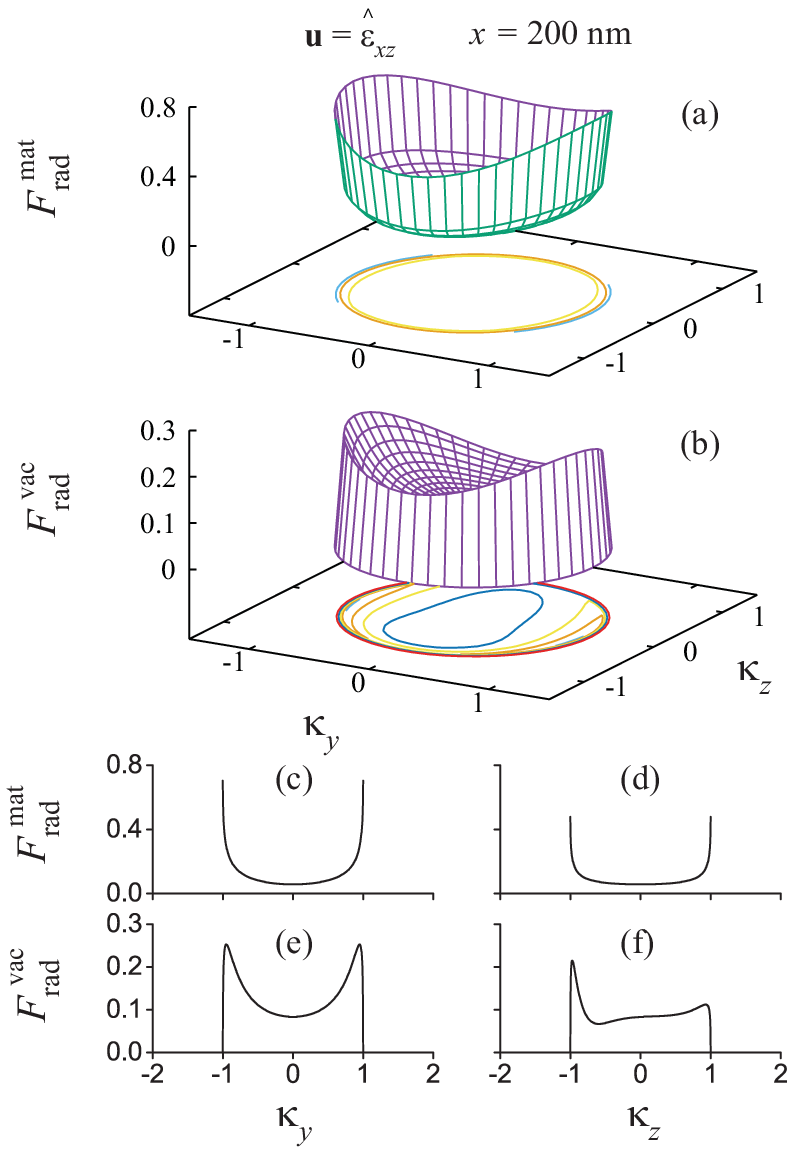}
\end{center}
\caption{(Color online) 
Angular densities $F_{\mathrm{rad}}^{\mathrm{mat}}$ (a) and  $F_{\mathrm{rad}}^{\mathrm{vac}}$ (b) of the rates of spontaneous emission into radiation modes
with the outputs inside and outside the dielectric, respectively, in the case of Fig.~\ref{fig6}, where the dipole polarization vector is $\mathbf{u}=\hat{\boldsymbol{\varepsilon}}_{xz}$ and the atom-interface distance is $x=200$~nm. Other parameters are as for Fig.~\ref{fig2}. The bottom panel shows the one-dimensional profiles of $F_{\mathrm{rad}}^{\mathrm{mat}}$ and $F_{\mathrm{rad}}^{\mathrm{vac}}$. In (c) and (e), $\kappa_z=0$. In (d) and (f), $\kappa_y=0$. 
}
\label{fig12}
\end{figure}

\begin{figure}[tbh]
\begin{center}
\includegraphics{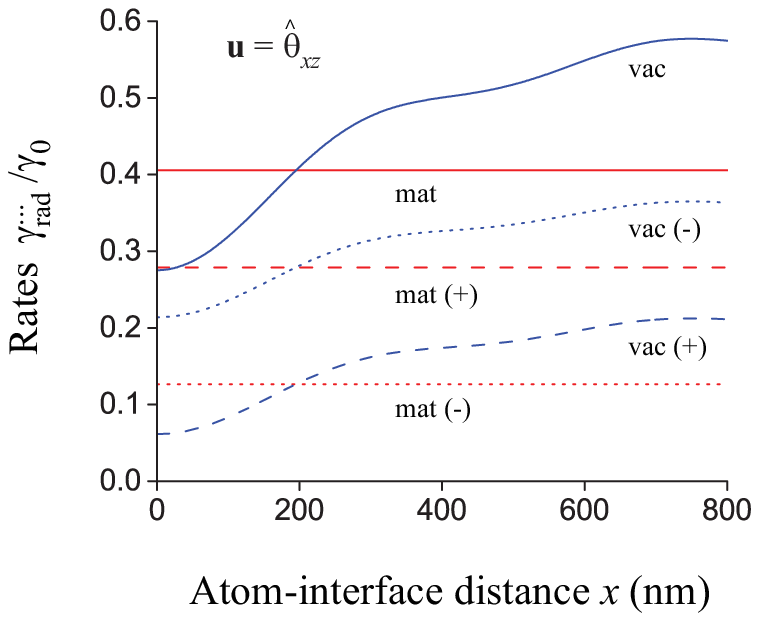}
\end{center}
\caption{(Color online) Rate  $\gamma_{\mathrm{rad}}^{\mathrm{mat}}$ and its components $\gamma_{\mathrm{rad}}^{\mathrm{mat}(\pm)}$
for radiation modes with outputs in the dielectric (red curves) and rate $\gamma_{\mathrm{rad}}^{\mathrm{vac}}$
and its components $\gamma_{\mathrm{rad}}^{\mathrm{vac}(\pm)}$ for radiation modes with outputs in the vacuum  (blue curves)
as functions of the atom-interface distance $x$. The polarization vector of the atomic dipole is $\mathbf{u}=\hat{\boldsymbol{\theta}}_{xz}$.
The rates are normalized to the spontaneous emission rate $\gamma_0$ of the atom in free space.
Other parameters are as for Fig.~\ref{fig2}.
}
\label{fig13}
\end{figure}

\begin{figure}[tbh]
\begin{center}
\includegraphics{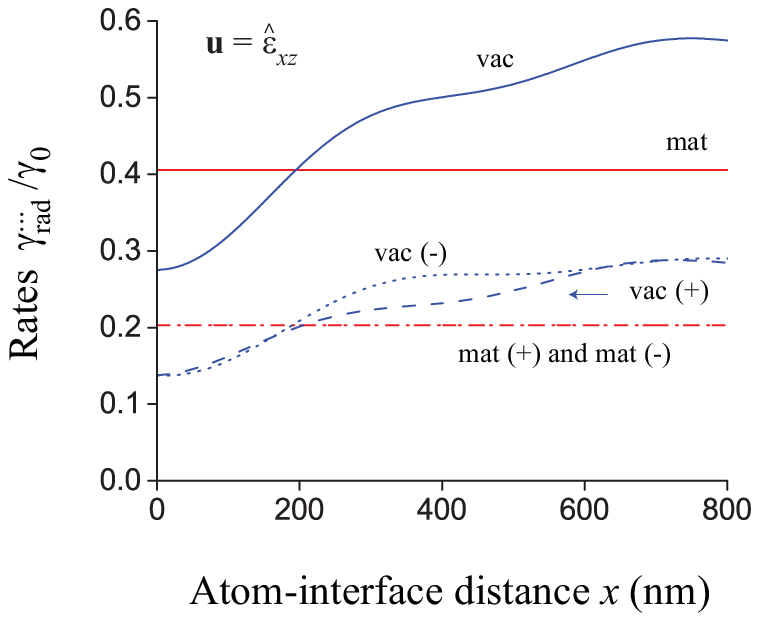}
\end{center}
\caption{(Color online) 
Rate  $\gamma_{\mathrm{rad}}^{\mathrm{mat}}$ and its components $\gamma_{\mathrm{rad}}^{\mathrm{mat}(\pm)}$
for radiation modes with outputs in the dielectric (red curves) and rate $\gamma_{\mathrm{rad}}^{\mathrm{vac}}$
and its components $\gamma_{\mathrm{rad}}^{\mathrm{vac}(\pm)}$ for radiation modes with outputs in the vacuum (blue curves)
as functions of the atom-interface distance $x$. The polarization vector of the atomic dipole is $\mathbf{u}=\hat{\boldsymbol{\varepsilon}}_{xz}$.
The rates are normalized to the spontaneous emission rate $\gamma_0$ of the atom in free space.
Other parameters are as for Fig.~\ref{fig2}.
}
\label{fig14}
\end{figure}

According to the previous section, the angular densities $F_{\mathrm{rad}}^{\mathrm{mat}}$ and  $F_{\mathrm{rad}}^{\mathrm{vac}}$ of the rates of spontaneous emission into radiation modes
outgoing into the dielectric and the vacuum, respectively, are given by Eqs.~\eqref{fl62} and \eqref{fl63}, respectively. Unlike the angular densities 
$F_{\mathrm{evan}}$ and $F_{\mathrm{rad}}$, the dielectric-side component $F_{\mathrm{rad}}^{\mathrm{mat}}$ and the vacuum-side component $F_{\mathrm{rad}}^{\mathrm{vac}}$ of $F_{\mathrm{rad}}$
can be asymmetric with respect to central inversion in the interface plane when the dipole polarization vector $\mathbf{u}$ is a real vector tilted with respect to the axis $x$ and to the interface plane $yz$.
In order to get insight into the asymmetry of the angular densities $F_{\mathrm{rad}}^{\mathrm{mat}}$ and  $F_{\mathrm{rad}}^{\mathrm{vac}}$ with respect to central inversion in the interface plane,
we present in Figs.~\ref{fig11}--\ref{fig14} the results of numerical calculations for these distribution functions and their related rates
in the cases of Fig.~\ref{fig5}, where $\mathbf{u}=\hat{\boldsymbol{\theta}}_{xz}$, and Fig.~\ref{fig6}, where $\mathbf{u}=\hat{\boldsymbol{\varepsilon}}_{xz}$.  

We plot the angular densities $F_{\mathrm{rad}}^{\mathrm{mat}}$ and $F_{\mathrm{rad}}^{\mathrm{vac}}$ in Figs.~\ref{fig11} and \ref{fig12} for the cases of $\mathbf{u}=\hat{\boldsymbol{\theta}}_{xz}$ and $\mathbf{u}=\hat{\boldsymbol{\varepsilon}}_{xz}$, respectively. We observe from Fig.~\ref{fig11} that, in the case where $\mathbf{u}=\hat{\boldsymbol{\theta}}_{xz}$, both $F_{\mathrm{rad}}^{\mathrm{mat}}$ and  $F_{\mathrm{rad}}^{\mathrm{vac}}$ are asymmetric with respect to the transformation $\boldsymbol{\kappa}\to-\boldsymbol{\kappa}$. This asymmetry of $F_{\mathrm{rad}}^{\mathrm{mat}}$ and $F_{\mathrm{rad}}^{\mathrm{vac}}$ is a consequence of the asymmetry of the orientation of the dipole polarization vector $\mathbf{u}$ with respect to the interface. We note that the difference $F_{\mathrm{rad}}^{\mathrm{mat}}(\kappa_y,\kappa_z)-F_{\mathrm{rad}}^{\mathrm{mat}}(-\kappa_y,-\kappa_z)$, which characterizes the asymmetry of $F_{\mathrm{rad}}^{\mathrm{mat}}$, is exactly opposite to the difference $F_{\mathrm{rad}}^{\mathrm{vac}}(\kappa_y,\kappa_z)-F_{\mathrm{rad}}^{\mathrm{vac}}(-\kappa_y,-\kappa_z)$, which characterizes the asymmetry of $F_{\mathrm{rad}}^{\mathrm{vac}}$. Due to the cancellation of the asymmetry in the sum, the density $F_{\mathrm{rad}}=F_{\mathrm{rad}}^{\mathrm{mat}}+F_{\mathrm{rad}}^{\mathrm{vac}}$ is symmetric with respect to the transformation $\boldsymbol{\kappa}\to-\boldsymbol{\kappa}$  [see Figs.~\ref{fig5}(b), \ref{fig5}(e), and \ref{fig5}(f)]. Figure \ref{fig12} shows that, in the case where $\mathbf{u}=\hat{\boldsymbol{\varepsilon}}_{xz}$, the distribution $F_{\mathrm{rad}}^{\mathrm{mat}}$ [see Figs.~\ref{fig12}(a), \ref{fig12}(c), and \ref{fig12}(d)] is symmetric  and the distribution $F_{\mathrm{rad}}^{\mathrm{vac}}$ [see Figs.~\ref{fig12}(b), \ref{fig12}(e), and \ref{fig12}(f)] is asymmetric with respect to the transformation $\boldsymbol{\kappa}\to-\boldsymbol{\kappa}$. The asymmetry of $F_{\mathrm{rad}}^{\mathrm{vac}}$ in Fig.~\ref{fig12}
is a consequence of the overlap between the ellipticity vector of the atomic dipole polarization and the ellipticity vector of the field mode polarization.
The symmetry of $F_{\mathrm{rad}}^{\mathrm{mat}}$ in Fig.~\ref{fig12} is a consequence of the fact that we have $\mathrm{Re\,}(u_x^*u_y)=\mathrm{Re\,}(u_x^*u_z)=0$ in the case considered.
When $\mathrm{Re\,}(u_x^*u_y)$ or $\mathrm{Re\,}(u_x^*u_z)$ is not zero, $F_{\mathrm{rad}}^{\mathrm{mat}}$ is not symmetric with respect to the transformation $\boldsymbol{\kappa}\to-\boldsymbol{\kappa}$.

We plot in Figs.~\ref{fig13} and \ref{fig14} the 
rate $\gamma_{\mathrm{rad}}^{\mathrm{mat}}$ and its components $\gamma_{\mathrm{rad}}^{\mathrm{mat}(\pm)}$
for radiation modes with outputs in the dielectric (red curves) and the rate $\gamma_{\mathrm{rad}}^{\mathrm{vac}}$
and its components $\gamma_{\mathrm{rad}}^{\mathrm{vac}(\pm)}$ for radiation modes with outputs in the vacuum (blue curves)
as functions of the atom-interface distance $x$. The polarization vector of the atomic dipole is $\mathbf{u}=\hat{\boldsymbol{\theta}}_{xz}$ in the case of Fig.~\ref{fig13} and is 
$\mathbf{u}=\hat{\boldsymbol{\varepsilon}}_{xz}$ in the case of Fig.~\ref{fig14}. 
Figures \ref{fig13} and \ref{fig14} show that $\gamma_{\mathrm{rad}}^{\mathrm{mat}}$ and $\gamma_{\mathrm{rad}}^{\mathrm{mat}(\pm)}$ (red curves)
do not depend on the distance $x$ while $\gamma_{\mathrm{rad}}^{\mathrm{vac}}$ and $\gamma_{\mathrm{rad}}^{\mathrm{vac}(\pm)}$ (blue curves) vary non-monotonically with increasing $x$.

Comparison between Figs.~\ref{fig13} and \ref{fig14} shows that we obtain the same values for
$\gamma_{\mathrm{rad}}^{\mathrm{mat}}$ (solid red curves) and the same values for $\gamma_{\mathrm{rad}}^{\mathrm{vac}}$ (solid blue curves) in the two cases. 
The reason is that the rates $\gamma_{\mathrm{rad}}^{\mathrm{mat}}=\gamma_{\mathrm{rad}}^{\mathrm{mat}(+)}+\gamma_{\mathrm{rad}}^{\mathrm{mat}(-)}$ and $\gamma_{\mathrm{rad}}^{\mathrm{vac}}=\gamma_{\mathrm{rad}}^{\mathrm{vac}(+)}+\gamma_{\mathrm{rad}}^{\mathrm{vac}(-)}$ depend on $|u_x|^2$ but not on the cross terms of the type $u^*_{j} u_{j'}$ where $j\not=j'$
and $j,j'=x,y,z$ [see Eqs.~\eqref{fl64} and \eqref{fl65}]. We note the following interesting features: 
$\gamma_{\mathrm{rad}}^{\mathrm{mat}}\simeq 0.4\gamma_0<\gamma_0/2$,
$\gamma_{\mathrm{rad}}^{\mathrm{vac}}<\gamma_{\mathrm{rad}}^{\mathrm{mat}}$ and 
$\gamma_{\mathrm{rad}}^{\mathrm{vac}}>\gamma_{\mathrm{rad}}^{\mathrm{mat}}$ for $x<195$ nm and $x>195$ nm, respectively,
$\gamma_{\mathrm{rad}}^{\mathrm{vac}}>\gamma_0/2$ for $x>397$ nm, 
and $\gamma_{\mathrm{rad}}^{\mathrm{vac}}$ tends to approach the limiting value $1-\gamma_{\mathrm{rad}}^{\mathrm{mat}}\sim 0.6\gamma_0$ in the limit $x\to+\infty$.

Figure \ref{fig13} shows that, in the case where $\mathbf{u}=\hat{\boldsymbol{\theta}}_{xz}$, the difference  
$\gamma_{\mathrm{rad}}^{\mathrm{mat}(+)}-\gamma_{\mathrm{rad}}^{\mathrm{mat}(-)}$ (see the dashed and dotted red curves) is a nonzero constant and is equal to the difference $\gamma_{\mathrm{rad}}^{\mathrm{vac}(-)}-\gamma_{\mathrm{rad}}^{\mathrm{vac}(+)}$ (see the dotted and dashed blue curves). This difference is caused by the tilting of the dipole polarization vector $\mathbf{u}$ with respect to the axis $x$ and the interface plane $yz$ [see expression \eqref{fl71} and the first term in expression \eqref{fl72}].

Figure \ref{fig14} shows that, in the case where $\mathbf{u}=\hat{\boldsymbol{\varepsilon}}_{xz}$, we have $\gamma_{\mathrm{rad}}^{\mathrm{mat}(+)}=\gamma_{\mathrm{rad}}^{\mathrm{mat}(-)}$ (see the dash-dotted red curve) and $\gamma_{\mathrm{rad}}^{\mathrm{vac}(+)}\not=\gamma_{\mathrm{rad}}^{\mathrm{vac}(-)}$ (see the dashed and dotted blue curves). 
The difference $\gamma_{\mathrm{rad}}^{\mathrm{vac}(+)}-\gamma_{\mathrm{rad}}^{\mathrm{vac}(-)}$ can be positive or negative depending on the distance $x$.
This difference is caused by spin-orbit coupling of light [see the second term in expression \eqref{fl72} and Eqs.~\eqref{fl45}--\eqref{fl47}].

\begin{figure}[tbh]
\begin{center}
\includegraphics{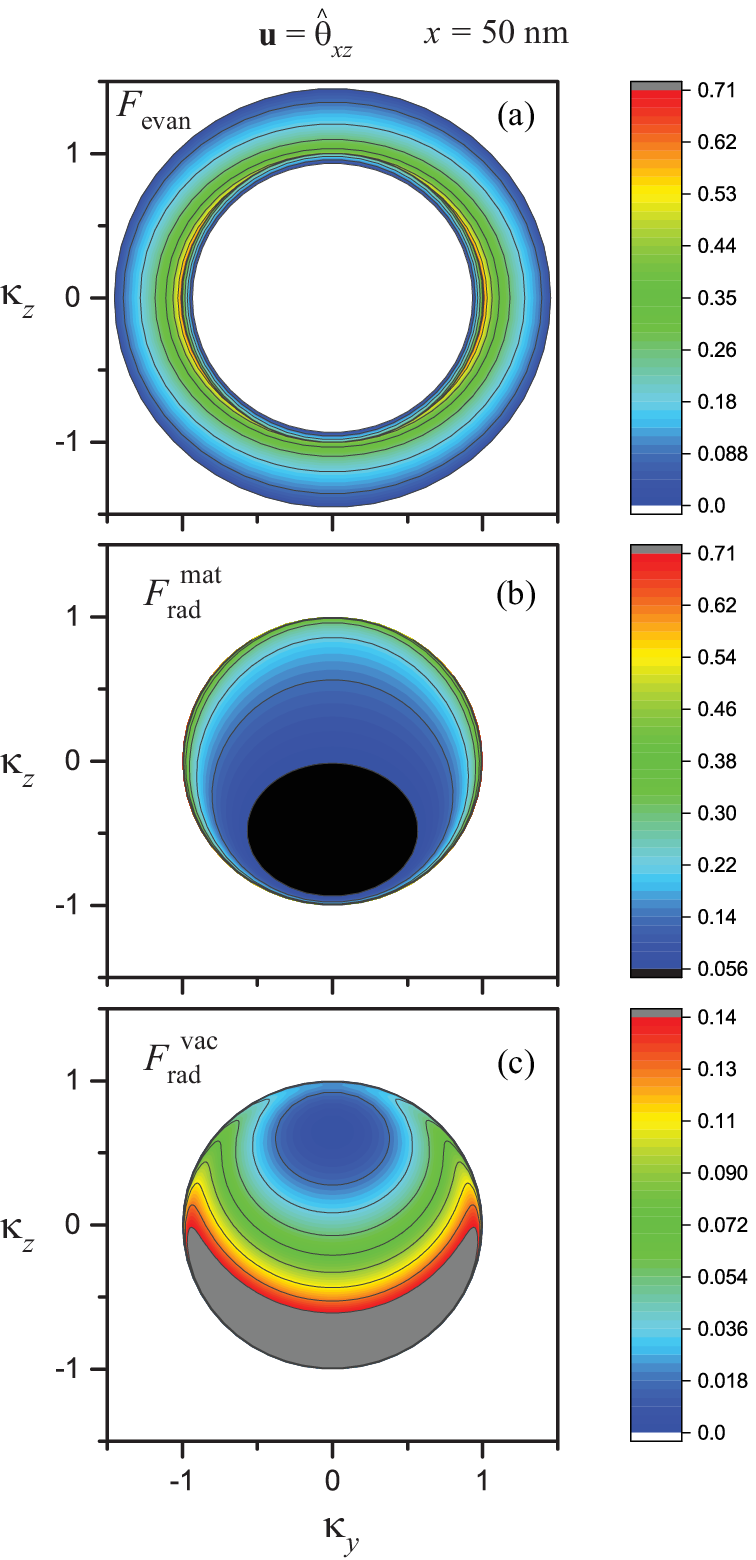}
\end{center}
\caption{(Color online) Angular densities $F_{\mathrm{evan}}$ (a), $F_{\mathrm{rad}}^{\mathrm{mat}}$ (b), and $F_{\mathrm{rad}}^{\mathrm{vac}}$ (c) 
of the rates of spontaneous emission into evanescent modes, radiation modes with outputs in the dielectric, and radiation modes with outputs in the vacuum, respectively,
in the case where the dipole polarization vector is $\mathbf{u}=\hat{\boldsymbol{\theta}}_{xz}$ and the atom-interface distance is $x=50$ nm. Other parameters are as for Fig.~\ref{fig2}.
}
\label{fig15}
\end{figure}

\begin{figure}[tbh]
\begin{center}
\includegraphics{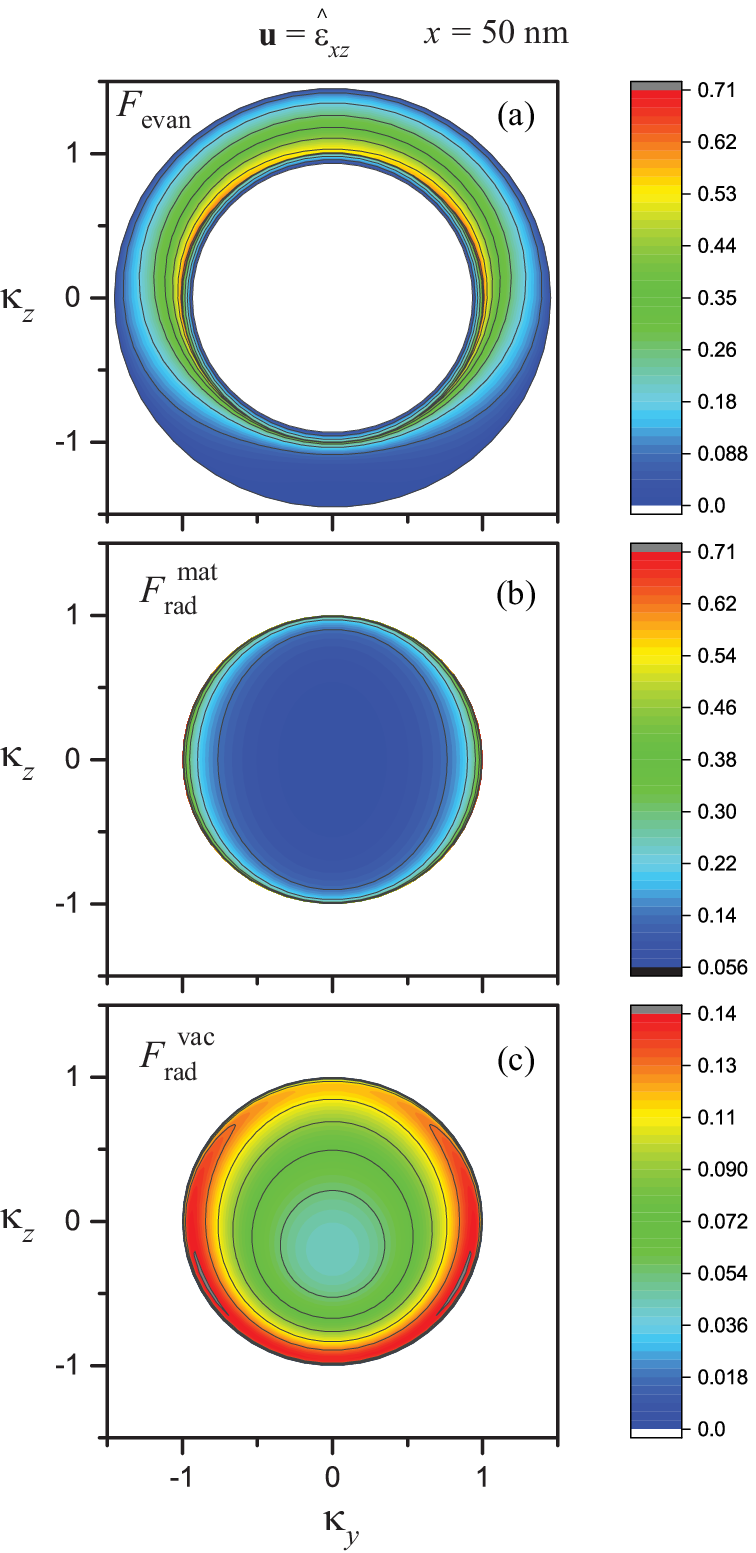}
\end{center}
\caption{(Color online) Angular densities $F_{\mathrm{evan}}$ (a), $F_{\mathrm{rad}}^{\mathrm{mat}}$ (b), and $F_{\mathrm{rad}}^{\mathrm{vac}}$ (c) 
of the rates of spontaneous emission into evanescent modes, radiation modes with outputs in the dielectric, and radiation modes with outputs in the vacuum, respectively,
in the case where the dipole polarization vector is $\mathbf{u}=\hat{\boldsymbol{\varepsilon}}_{xz}$ and the atom-interface distance is $x=50$ nm. Other parameters are as for Fig.~\ref{fig2}.
}
\label{fig16}
\end{figure}

The angular distributions of the rates of emission of a dipole-like particle can be measured experimentally by direct imaging the emission patterns in the back focal plane of a high-numerical-aperture objective lens \cite{Leuchs2014,Lieb2004,Banzer2010}. The images are the contour plots of the angular densities of the rates of emission. We show the color-filled contour plots of the angular densities $F_{\mathrm{evan}}$, $F_{\mathrm{rad}}^{\mathrm{mat}}$, and $F_{\mathrm{rad}}^{\mathrm{vac}}$ in Figs.~\ref{fig15} and \ref{fig16} for the cases where 
$\mathbf{u}=\hat{\boldsymbol{\theta}}_{xz}$ and $\mathbf{u}=\hat{\boldsymbol{\varepsilon}}_{xz}$, respectively. The atom-interface distance is chosen to be $x=50$ nm. 
Figure~\ref{fig15} shows that, in the case of $\mathbf{u}=\hat{\boldsymbol{\theta}}_{xz}$, 
the function $F_{\mathrm{evan}}$ [see Fig.~\ref{fig15}(a)] is symmetric but the functions $F_{\mathrm{rad}}^{\mathrm{mat}}$ [see Fig.~\ref{fig15}(b)] and $F_{\mathrm{rad}}^{\mathrm{vac}}$ [see Fig.~\ref{fig15}(c)] are not symmetric with respect to central inversion in the interface plane.
Figure~\ref{fig16} shows that, in the case of $\mathbf{u}=\hat{\boldsymbol{\varepsilon}}_{xz}$, the function $F_{\mathrm{rad}}^{\mathrm{mat}}$ [see Fig.~\ref{fig16}(b)] is symmetric but the functions $F_{\mathrm{evan}}$ [see Fig.~\ref{fig16}(a)] and $F_{\mathrm{rad}}^{\mathrm{vac}}$ [see Fig.~\ref{fig16}(c)] are not symmetric with respect to central inversion in the interface plane.

\begin{figure}[tbh]
\begin{center}
\includegraphics{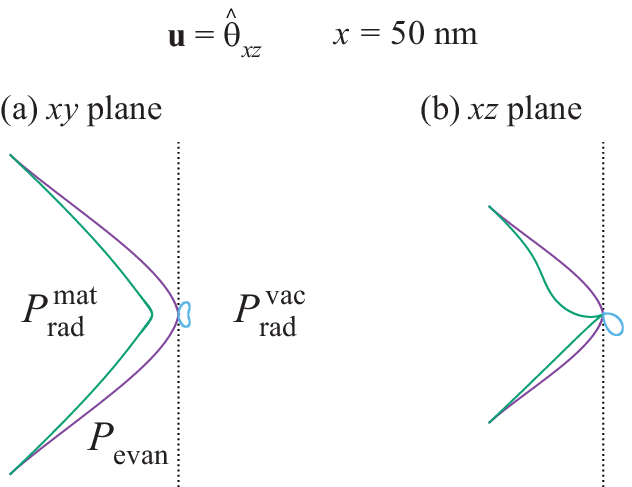}
\end{center}
\caption{(Color online) Radiation patterns $P_{\mathrm{evan}}$ (blue curves), $P_{\mathrm{rad}}^{\mathrm{mat}}$ (green curves), and $P_{\mathrm{rad}}^{\mathrm{vac}}$ (cyan curves) 
for evanescent modes, radiation modes with outputs in the dielectric, and radiation modes with outputs in the vacuum, respectively,
in the case where the dipole polarization vector is $\mathbf{u}=\hat{\boldsymbol{\theta}}_{xz}$ and the atom-interface distance is $x=50$ nm. The horizontal axis of the figure is the direction of the $x$ axis. In (a), we set $\phi=0,\pi$ to calculate the patterns in the $xy$ plane. In (b), we set $\phi=\pm \pi/2$ to calculate the patterns in the $xz$ plane. Other parameters are as for Fig.~\ref{fig2}.
}
\label{fig17}
\end{figure}

\begin{figure}[tbh]
\begin{center}
\includegraphics{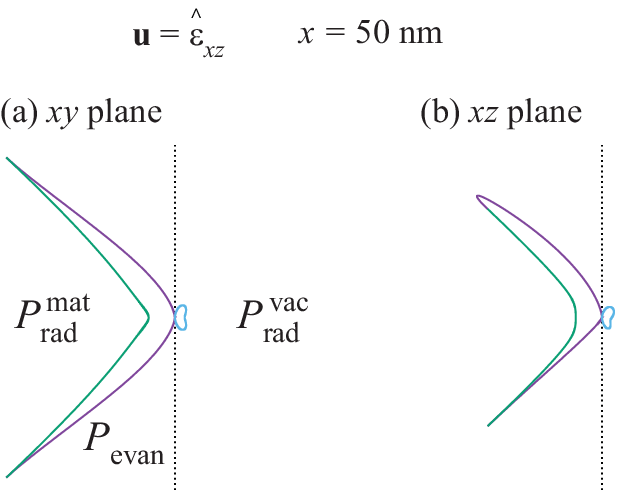}
\end{center}
\caption{(Color online) Radiation patterns $P_{\mathrm{evan}}$ (blue curves), $P_{\mathrm{rad}}^{\mathrm{mat}}$  (green curves), and $P_{\mathrm{rad}}^{\mathrm{vac}}$ (cyan curves) 
for evanescent modes, radiation modes with outputs in the dielectric, and radiation modes with outputs in the vacuum, respectively,
in the case where the dipole polarization vector is $\mathbf{u}=\hat{\boldsymbol{\varepsilon}}_{xz}$ and the atom-interface distance is $x=50$ nm. The horizontal axis of the figure is the direction of the $x$ axis. In (a), we set $\phi=0,\pi$ to calculate the patterns in the $xy$ plane. In (b), we set $\phi=\pm \pi/2$ to calculate the patterns in the $xz$ plane. Other parameters are as for Fig.~\ref{fig2}.
}
\label{fig18}
\end{figure}

In the far-field limit, the radiation patterns of emission into evanescent modes, radiation modes with outputs in the dielectric, and radiation modes with outputs in the vacuum are described by the functions $P_{\mathrm{evan}}(\theta,\phi)$, $P_{\mathrm{rad}}^{\mathrm{mat}}(\theta,\phi)$, and $P_{\mathrm{rad}}^{\mathrm{vac}}(\theta,\phi)$, respectively, 
We plot these functions in Figs.~\ref{fig17} and \ref{fig18} for the cases where 
$\mathbf{u}=\hat{\boldsymbol{\theta}}_{xz}$ and $\mathbf{u}=\hat{\boldsymbol{\varepsilon}}_{xz}$, respectively. The atom-interface distance is chosen to be $x=50$ nm. The horizontal axis of the figures is the direction of the $x$ axis. Figures \ref{fig17}(a) and \ref{fig18}(a) show that the radiation patterns in the $xy$ plane are symmetric with respect to the $x$ axis.
We observe from Fig.~\ref{fig17}(b) that, in the case where $\mathbf{u}=\hat{\boldsymbol{\theta}}_{xz}$, 
the pattern $P_{\mathrm{evan}}$ in the $xz$ plane is symmetric with respect to the $x$ axis but
the patterns $P_{\mathrm{rad}}^{\mathrm{mat}}$ and $P_{\mathrm{rad}}^{\mathrm{vac}}$ are not. 
Figure \ref{fig18}(b) shows that, in the case where $\mathbf{u}=\hat{\boldsymbol{\varepsilon}}_{xz}$, 
the pattern $P_{\mathrm{rad}}^{\mathrm{mat}}$ in the $xz$ plane is symmetric with respect to the $x$ axis but
the patterns $P_{\mathrm{evan}}$ and $P_{\mathrm{rad}}^{\mathrm{vac}}$ are not. These features are in agreement with the analytical results presented in the previous section.

\section{Summary}
\label{sec:summary}

We have studied spontaneous emission of a two-level atom with an arbitrarily polarized electric dipole in front of a flat dielectric surface. 
We have treated the general case where the atomic dipole matrix element is a complex vector, that is, the atomic dipole can rotate with time in space.
In order to get deep insight into the underlying physics, we have employed a full quantum formalism for the atom and the field, and have used the Hamiltonian method and the mode expansion approach.
We have calculated the rates of spontaneous emission into evanescent and radiation modes. We have examined the angular densities of the rates of spontaneous emission in the space of wave vectors for the field modes. We have found that, when the ellipticity of the atomic dipole is not zero, the angular density of the spontaneous emission rate of the atom may have different values for the modes with the opposite in-plane (transverse) wave vectors. We have shown that the asymmetry of the angular density of the spontaneous emission rate under central inversion in the space of transverse wave vectors is a result of spin-orbit coupling of light and occurs when the ellipticity vector of the atomic dipole polarization overlaps with the ellipticity vector of the field mode polarization.  

Since the ellipticity of the electric polarization of the TE modes is zero, only the TM modes can contribute to the asymmetry of spontaneous emission with respect to central inversion
in the interface plane. The ellipticity of the electric polarization of the TM evanescent mode $(\omega\mathbf{K}p1)$ arises as a consequence of the fact that the field in this evanescent mode has a longitudinal component whose phase is shifted by $\pi/2$ from that of the transverse component. Due to the fast decay of the field in the evanescent modes, the difference between the rates of spontaneous emission into evanescent modes with opposite in-plane wave vectors decreases monotonically with increasing distance from the atom to the interface. This difference achieves its maximum value when the atom is positioned on the surface of the dielectric. Meanwhile, the ellipticity of the electric polarization of the TM radiation mode $(\omega\mathbf{K}p2)$ results from the interference between the incident and reflected fields in this mode, which have different polarization vectors and different phases. Due to the oscillatory behavior of interference, the difference between the rates of spontaneous emission into radiation modes with opposite in-plane wave vectors oscillates with increasing distance from the atom to the interface. This difference can be positive or negative depending on the atom-interface distance $x$, and is zero for $x=0$. The lack of asymmetry for radiation modes under the in-plane central inversion in the case of $x=0$ is a consequence of the fact that the relative phase between the incident and reflected fields at $x=0$ is just the phase of the reflection coefficient and hence is equal to $0$ or $\pi$.

We have shown that the ellipticity of the atomic dipole affects the angular density of the rate of spontaneous emission into the radiation modes outgoing into the vacuum. However, this ellipticity does not modify the angular density of the rate of spontaneous emission into the radiation modes outgoing into the dielectric.

The results of this paper can be used not only for spontaneous emission of a two-level atom with an arbitrarily polarized dipole but also for the rate enhancement factor and the radiation pattern of an arbitrarily polarized classical oscillating dipole.
These results can also be extended to the case of a multilevel atom by summing up the contributions from different transitions from each upper level. Due to the competition between different types of transitions, the directional dependence of the spontaneous emission rate of a multilevel atom is, in general, weaker than that of a two-level atom with a circularly polarized dipole.

\begin{acknowledgments}
F.L.K. acknowledges support by the European Commission (Marie Curie IIF Grant No. 332255). 
\end{acknowledgments}

\appendix

\section{Tensor decomposition}
\label{sec:decomposition}

We use the Cartesian coordinate frame $\{x,y,z\}$.
The spherical tensor components $A_q$, with $q=-1,0,1$, of an arbitrary complex vector $\mathbf{A}=\{A_x,A_y,A_z\}$ are given by
\begin{eqnarray}\label{h3} 
A_{-1}&=&(A_{x}-iA_{y})/\sqrt{2}, \nonumber\\ 
A_0&=& A_{z}, \nonumber\\ 
A_{1}&=& -(A_{x}+iA_{y})/\sqrt{2}.
\end{eqnarray}
The absolute length of the complex vector $\mathbf{A}$ is given by $|\mathbf{A}|=\sqrt{|A_x|^2+|A_y|^2+|A_z|^2}$.
The compound tensor components $\{\mathbf{A}^*\otimes\mathbf{A}\}_{Kq}$, where $K=0,1,2$ and $q=-K,\dots,K$, are given by
\begin{equation}\label{h4}
\{\mathbf{A}^*\otimes\mathbf{A}\}_{0,0}=-\frac{|A_0|^2+|A_1|^2+|A_{-1}|^2}{\sqrt3},
\end{equation}
\begin{eqnarray}\label{h5}
\{\mathbf{A}^*\otimes\mathbf{A}\}_{1,0}&=&\frac{|A_1|^2-|A_{-1}|^2}{\sqrt2},
\nonumber\\
\{\mathbf{A}^*\otimes\mathbf{A}\}_{1,1}&=&-\frac{A_0A^*_{-1}+A_0^*A_1}{\sqrt2},
\nonumber\\
\{\mathbf{A}^*\otimes\mathbf{A}\}_{1,-1}&=&\frac{A_0A^*_1+A_0^*A_{-1}}{\sqrt2},
\end{eqnarray}
and
\begin{eqnarray}\label{h6}
\{\mathbf{A}^*\otimes\mathbf{A}\}_{2,0}&=&\frac{2|A_{0}|^2-|A_1|^2-|A_{-1}|^2}{\sqrt6},
\nonumber\\
\{\mathbf{A}^*\otimes\mathbf{A}\}_{2,1}&=&-\frac{A_0A^*_{-1}-A^*_{0}A_1}{\sqrt2},
\nonumber\\
\{\mathbf{A}^*\otimes\mathbf{A}\}_{2,-1}&=&-\frac{A_0A^*_{1}-A^*_{0}A_{-1}
}{\sqrt2},
\nonumber\\
\{\mathbf{A}^*\otimes\mathbf{A}\}_{2,2}&=&-A_{1}A^*_{-1},
\nonumber\\
\{\mathbf{A}^*\otimes\mathbf{A}\}_{2,-2}&=&-A_{-1}A^*_{1}.
\end{eqnarray}

The scalar product of arbitrary complex vectors $\mathbf{A}$ and $\mathbf{B}$ is defined by 
\begin{equation}\label{h7}
\mathbf{A}\cdot\mathbf{B}=A_xB_x+A_yB_y+A_zB_z=\sum_{q=-1}^{1}(-1)^qA_qB_{-q}. 
\end{equation}
We have the relation $|\mathbf{A}|^2=\mathbf{A}^*\cdot\mathbf{A}$.
The vector product of the vectors $\mathbf{A}$ and $\mathbf{B}$ is defined by
\begin{equation}\label{h8}
\begin{split}
[\mathbf{A}\times\mathbf{B}]&=(A_yB_z-A_zB_y)\hat{\mathbf{x}}+(A_zB_x-A_xB_z)\hat{\mathbf{y}}\\
&\quad + (A_xB_y-A_yB_x)\hat{\mathbf{z}}. 
\end{split}
\end{equation}
According to \cite{tensor books}, we have
\begin{equation}\label{h10}
|\mathbf{A}\cdot\mathbf{B}|^2=\sum_{K=0,1,2} (-1)^K \{\mathbf{A}^*\otimes\mathbf{A}\}_K\cdot\{\mathbf{B}^*\otimes\mathbf{B}\}_K.
\end{equation}
The above formula can be rewritten in the form
\begin{equation}\label{h11}
\begin{split}
|\mathbf{A}\cdot\mathbf{B}|^2&=\frac{1}{3}|\mathbf{A}|^2|\mathbf{B}|^2+\frac{1}{2}[\mathbf{A}^*\times\mathbf{A}]\cdot[\mathbf{B}^*\times\mathbf{B}]\\
&\quad + \{\mathbf{A}^*\otimes\mathbf{A}\}_2\cdot\{\mathbf{B}^*\otimes\mathbf{B}\}_2.
\end{split}
\end{equation}


\end{document}